\documentclass[11pt,english]{article}
\usepackage{lmodern}
\usepackage{lmodern}
\usepackage[T1]{fontenc}
\usepackage[latin9]{inputenc}
\usepackage{xcolor}
\usepackage{booktabs}
\usepackage{amsmath}
\usepackage{amssymb}
\usepackage{cancel}
\usepackage{stackrel}
\usepackage{graphicx}
\usepackage{setspace}
\usepackage{wasysym}
\makeatletter

\providecommand{\tabularnewline}{\\}

\usepackage{jheppub}

\usepackage{subfig}
\usepackage[compat=1.1.0]{tikz-feynman}
\usepackage[normalem]{ulem}
\usepackage{geometry}
\usepackage{booktabs}
\hypersetup{
linkcolor=magenta,
urlcolor=magenta,
citecolor=magenta,
linktocpage=true
}

\title{\boldmath Tri-hypercharge: a separate gauged weak hypercharge for each fermion family as the origin of flavour}

\author[]{Mario Fern\'andez Navarro}
\author[]{and Stephen F. King}
\affiliation[]{School of Physics \& Astronomy, University of Southampton, Southampton SO17 1BJ, UK}

\emailAdd{M.F.Navarro@soton.ac.uk}
\emailAdd{S.F.King@soton.ac.uk}

\abstract{We propose a tri-hypercharge (TH) embedding of the Standard Model (SM) in which a separate gauged weak hypercharge is associated with each
fermion family. In this way, every quark and lepton multiplet carries unique gauge quantum numbers under the extended gauge group,
providing the starting point for a theory of flavour.
If the Higgs doublets only carry third family hypercharge, then only third family renormalisable Yukawa couplings are allowed.
However, non-renormalisable Yukawa couplings may be induced by the high scale Higgs fields (hyperons) which break the three hypercharges down
to the SM hypercharge, providing an explanation for fermion mass hierarchies and 
the smallness of CKM quark mixing. Following a similar methodology,
we study the origin of neutrino masses and mixing in this model. Due
to the TH gauge symmetry, the implementation of a seesaw
mechanism naturally leads to a low scale seesaw, where the right-handed
neutrinos in the model may be as light as the TeV scale. 
We present simple examples of hyperon fields which can reproduce all quark and lepton (including neutrino) masses and mixing.
After a preliminary
phenomenological study, we conclude that one of the massive $Z'$
bosons can be as light as a few TeV, with implications for flavour-violating
observables, LHC physics and electroweak precision observables.}

\makeatother

\usepackage{babel}
\begin{document}
\makeatletter
\gdef\@fpheader{}
\makeatother

\maketitle \flushbottom

\pagenumbering{arabic}
\setcounter{page}{1}
\allowdisplaybreaks

\section{Introduction}

The existence of the three families of quarks and leptons is one of
the fundamental mysteries of the Standard Model (SM). Indeed many
of the parameters of the SM are associated with the resulting quark
and lepton mass patterns and mixings. The low energy masses of quarks
and charged leptons may be expressed approximately as \cite{PDG:2022ynf}
\begin{alignat}{2}
 & m_{t}\sim\frac{v_{\mathrm{SM}}}{\sqrt{2}}\,,\qquad & m_{c}\sim\lambda^{3.3}\frac{v_{\mathrm{SM}}}{\sqrt{2}}\,,\qquad & m_{u}\sim\lambda^{7.5}\frac{v_{\mathrm{SM}}}{\sqrt{2}}\,,\\
 & m_{b}\sim\lambda^{2.5}\frac{v_{\mathrm{SM}}}{\sqrt{2}}\,,\qquad & m_{s}\sim\lambda^{5.0}\frac{v_{\mathrm{SM}}}{\sqrt{2}}\,,\qquad & m_{d}\sim\lambda^{7.0}\frac{v_{\mathrm{SM}}}{\sqrt{2}}\,,\\
 & m_{\tau}\sim\lambda^{3.0}\frac{v_{\mathrm{SM}}}{\sqrt{2}}\,,\qquad & m_{\mu}\sim\lambda^{4.9}\frac{v_{\mathrm{SM}}}{\sqrt{2}}\,,\qquad & m_{e}\sim\lambda^{8.4}\frac{v_{\mathrm{SM}}}{\sqrt{2}}\,,
\end{alignat}
where $v_{\mathrm{SM}}\simeq246\,\mathrm{GeV}$ is the SM vacuum expectation
value (VEV) and $\lambda\simeq0.224$ is the Wolfenstein parameter
which parameterises the CKM matrix as
\begin{alignat}{2}
 & V_{us}\sim\lambda\,,\qquad & V_{cb}\sim\lambda^{2}\,,\qquad & V_{ub}\sim\lambda^{3}\,.
\end{alignat}
The hierarchical patterns of masses and CKM mixing, and the possibility
that they might be understood in the form of a more fundamental \textit{theory
of flavour} beyond the SM, has classically been denoted as the \textit{flavour
puzzle}. The discovery of neutrino oscillations, which proves that
at least two neutrinos have non-zero mass, has made the flavour puzzle
difficult to ignore, enlarging the flavour sector with extra neutrino
mixing angles \cite{deSalas:2020pgw,Gonzalez-Garcia:2021dve}
\begin{alignat}{2}
 & \tan\theta_{23}\sim1\,,\qquad & \tan\theta_{12}\sim\frac{1}{\sqrt{2}}\,,\qquad & \theta_{13}\sim\frac{\lambda}{\sqrt{2}}\,,
\end{alignat}
plus very tiny neutrino masses which follow either a normal or inverted
ordering. The need for a theory of flavour to understand all fermion
masses and mixings is now more urgent than ever.

Theories of flavour may involve new symmeties structures (global,
local, continuous, discrete, abelian, non-abelian...) beyond the SM
group, possibly broken at some high scale down to the SM. Traditionally,
new gauge structures beyond the SM have been considered to be flavour
universal, as grand unified theories (GUTs) which usually embed all
three families in an identical way, or even extended GUTs which embed
all three families in a single representation (usually along with
extra exotic fermions). Alternatively, there exists the well-motivated
case of family symmetries which commute with the SM gauge group, and
are then spontaneously broken, leading to family structure. However,
there are other less explored ways in which the SM gauge group could
be embedded into a larger gauge structure in a flavour non-universal
way. In particular, the \textit{family decomposition} of the SM gauge
group (including a hierarchical symmetry breaking pattern down to
the SM) was first proposed during the 80s and 90s, with the purpose
of motivating lepton non-universality \cite{Li:1981nk,Ma:1987ds,Ma:1988dn,Li:1992fi}
or assisting technicolor model building \cite{Hill:1994hp,Muller:1996dj,Malkawi:1996fs}.
However, the natural origin of flavour hierarchies in such a framework
was not explored until more recently in \cite{Craig:2011yk,Panico:2016ull,Barbieri:2021wrc}.
Here it was proposed that the flavour non-universality of Yukawa couplings
in the SM might well find its origin in a flavour non-universal gauge
sector, broken in a hierarchical way down to the SM. Interestingly,
model building in this direction has received particular attention
in recent years \cite{Bordone:2017bld,Allwicher:2020esa,Fuentes-Martin:2020pww,Fuentes-Martin:2022xnb,Davighi:2022bqf,Davighi:2022fer}.
With the exception of Ref.~\cite{Davighi:2022fer}, the remaining
recent attempts have been motivated by the need to obtain a TeV scale
vector leptoquark from Pati-Salam unification in order to address
the so-called $B$-anomalies\footnote{Remarkably, flavour non-universality is not the only way to connect
the TeV-scale Pati-Salam vector leptoquark addressing the $B$-anomalies
with the origin of flavour hierarchies, see the twin Pati-Salam theory
of flavour in \cite{King:2021jeo,FernandezNavarro:2022gst,FernandezNavarro:2023lgk,FernandezNavarro:2023ykw}
which considers the mechanism of messenger dominance \cite{Ferretti:2006df}.}. Therefore, all these setups share a similar feature: a low scale
$SU(4)$ gauge group under which only the third family of SM fermions
transforms in a non-trivial way. In contrast, in this work we want
to explore the capabilities of flavour non-universality to address
the flavour puzzle in a more minimal, simple and bottom-up approach.

We propose that the SM symmetry originates from a larger gauge group
in the UV that contains three separate weak hypercharge gauge factors,
\begin{equation}
SU(3)_{c}\times SU(2)_{L}\times U(1)_{Y_{1}}\times U(1)_{Y_{2}}\times U(1)_{Y_{3}}\,,
\end{equation}
which we will denote as the \textit{tri-hypercharge} (TH) $U(1)_{Y}^{3}$
gauge group. We will associate each of the three hypercharge gauge
groups with a separate SM family, such that each fermion family $i$
only carries hypercharge under the corresponding $U(1)_{Y_{i}}$ factor.
This ensures that each family transforms differently under the gauge
group $U(1)_{Y}^{3}$, which avoids the family repetition of the SM,
and provides the starting point for a theory of flavour. For example,
assuming that a single Higgs doublet only carries third family hypercharge,
then only the third family Yukawa couplings are allowed at renormalisable
level. With two Higgs doublets carrying third family hypercharge,
we show that the naturalness of the scheme increases. This simple
and economical framework naturally explains the heaviness of the third
family, the smallness of $V_{cb}$ and $V_{ub}$, and delivers Yukawa
couplings that preserve an accidental and global $U(2)^{5}$ flavour
symmetry acting on the light families, which is known to provide a
good first order description of the SM spectrum plus an efficient
suppression of flavour-violating effects for new physics \cite{Barbieri:2011ci}.
Remarkably, this appears to be the simplest way to provide the $U(2)^{5}$
flavour symmetry\footnote{An alternative way to deliver $U(2)^{5}$ consists of decomposing
$SU(2)_{L}$ only and taking advantage of the fact that right-handed
rotations remain unphysical in order to remove the remaining $U(2)^{5}$-breaking
entries of the Yukawa matrices (see the complete review of Ref.~\cite{Davighi:2023iks}).
Another example \cite{Allanach:2018lvl} considered an extension of
the SM by a $U(1)_{Y_{3}}$ gauge group under which only third family
fermions (and the Higgs) are hypercharge-like charged, where $U(1)_{Y_{3}}$
commutes with SM hypercharge, leading to an accidental $U(2)^{5}$.}. The masses of first and second family fermions, along with the CKM
mixing, then appear as small breaking sources of $U(2)^{5}$ that
arise after the cascade spontaneous symmetry breaking of $U(1)_{Y}^{3}$
down to SM hypercharge, which can be parameterised in a model-independent
way in terms of spurions. In a realistic model, the spurions will
be realised by a choice of ``hyperon'' scalars which transform under
the different family hypercharge groups, breaking the tri-hypercharge
symmetry. We will motivate a specific symmetry breaking chain where
dynamics at a low scale, which could be as low as the TeV, explain
the flavour hierarchies $m_{2}/m_{3}$, while dynamics at a heavier
scale explain $m_{1}/m_{2}$. This symmetry breaking pattern will
sequentially recover the approximate flavour symmetry of the SM, and
provide a natural suppression of FCNCs for TeV new physics, while
the rest of flavour-violating effects are suppressed by a naturally
heavier scale.

The paper is organised as follows. In Section~\ref{sec:The-Tri-Hypercharge-model}
we introduce the TH gauge group, along with the fermion and Higgs
doublet content of the model. We discuss the implications for third
family fermion masses along with the mass hierarchy between the top
and bottom/tau fermions. In Section~\ref{sec:Charged-fermion-masses-mixing}
we study the origin of charged fermion masses and mixing in the TH
model, firstly via a spurion formalism which reveals model-independent
considerations, and secondly by introducing example models with hyperons.
In Section~\ref{sec:Neutrino-masses-and-Mixing} we study the origin
of neutrino masses and mixing in the TH model. In particular, we discuss
the impact of the $U(2)^{5}$ flavour symmetry over the dimension-5
Weinberg operator, and afterwards we provide an example type I seesaw
model where neutrino masses and mixing can be accommodated. In Section~\ref{sec:Phenomenology}
we perform a preliminary exploration of the phenomenological implications
and discovery prospects of the $U(1)_{Y}^{3}$ theory of flavour.
Finally, Section~\ref{sec:Conclusions} outlines our main conclusions.\clearpage

\section{Tri-hypercharge gauge theory\label{sec:The-Tri-Hypercharge-model}}

\begin{figure}
\begin{centering}
\includegraphics[scale=0.4]{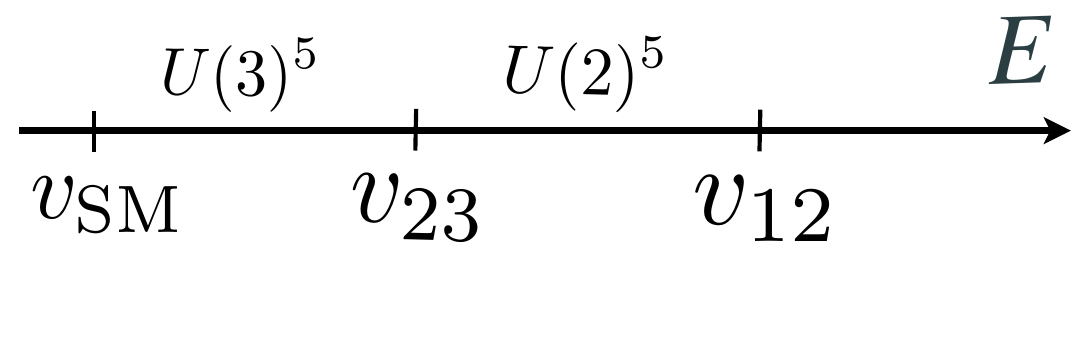} 
\par\end{centering}
\caption{Diagram showing the different energy scales in tri-hypercharge, along
with the approximate flavour symmetries that apply at each scale:
$v_{23}$ denotes the low scale where $U(2)^{5}$ is approximately
preserved, and the hierarchy $m_{2}/m_{3}$ is explained. $v_{12}$
denotes the higher scale where $U(2)^{5}$ is explicitly broken and
the hierarchy $m_{1}/m_{2}$ is explained.\label{fig:-Multiscale-picture-flavour}}
\end{figure}

The tri-hypercharge gauge group is based on assigning a separate gauged
weak hypercharge to each fermion family, 
\begin{table}[t]
\begin{centering}
\begin{tabular}{lccccc}
\toprule 
Field  & $SU(3)_{c}$  & $SU(2)_{L}$  & $U(1)_{Y_{1}}$  & $U(1)_{Y_{2}}$  & $U(1)_{Y_{3}}$\tabularnewline
\midrule
\midrule 
$Q_{1}$  & $\mathrm{\mathbf{3}}$  & $\mathbf{2}$  & $1/6$  & 0  & 0\tabularnewline
$u_{1}^{c}$  & $\bar{\mathbf{3}}$  & $\mathbf{1}$  & $-2/3$  & 0  & 0\tabularnewline
$d_{1}^{c}$  & $\bar{\mathbf{3}}$  & $\mathbf{1}$  & $1/3$  & 0  & 0\tabularnewline
$L_{1}$  & $\mathbf{1}$  & $\mathbf{2}$  & $-1/2$  & 0  & 0\tabularnewline
$e_{1}^{c}$  & $\mathbf{1}$  & $\mathbf{1}$  & $1$  & 0  & 0\tabularnewline
\midrule 
$Q_{2}$  & $\mathrm{\mathbf{3}}$  & $\mathbf{2}$  & 0  & $1/6$  & 0\tabularnewline
$u_{2}^{c}$  & $\bar{\mathbf{3}}$  & $\mathbf{1}$  & 0  & $-2/3$  & 0\tabularnewline
$d_{2}^{c}$  & $\bar{\mathbf{3}}$  & $\mathbf{1}$  & 0  & $1/3$  & 0\tabularnewline
$L_{2}$  & $\mathbf{1}$  & $\mathbf{2}$  & 0  & $-1/2$  & 0\tabularnewline
$e_{2}^{c}$  & $\mathbf{1}$  & $\mathbf{1}$  & 0  & $1$  & 0\tabularnewline
\midrule 
$Q_{3}$  & $\mathrm{\mathbf{3}}$  & $\mathbf{2}$  & 0  & 0  & $1/6$\tabularnewline
$u_{3}^{c}$  & $\bar{\mathbf{3}}$  & $\mathbf{1}$  & 0  & 0  & $-2/3$\tabularnewline
$d_{3}^{c}$  & $\bar{\mathbf{3}}$  & $\mathbf{1}$  & 0  & 0  & $1/3$\tabularnewline
$L_{3}$  & $\mathbf{1}$  & $\mathbf{2}$  & 0  & 0  & $-1/2$\tabularnewline
$e_{3}^{c}$  & $\mathbf{1}$  & $\mathbf{1}$  & 0  & 0  & $1$\tabularnewline
\bottomrule
\end{tabular}
\par\end{centering}
\caption{Charge assignments of the SM fermions under the TH gauge group. $Q_{i}$
and $L_{i}$ (where $i=1,2,3$) are left-handed (LH) $SU(2)_{L}$
doublets of chiral quarks and leptons, while $u_{i}^{c}$, $d_{i}^{c}$
and $e_{i}^{c}$ are the $CP$-conjugate right-handed (RH) quarks
and leptons (so that they become left-handed). In all cases we consider
a 2-component convention. \label{tab:Field_content}}
\end{table}
\begin{equation}
SU(3)_{c}\times SU(2)_{L}\times U(1)_{Y_{1}}\times U(1)_{Y_{2}}\times U(1)_{Y_{3}}\,,
\end{equation}
in such a way that the $i$th fermion family only carries $Y_{i}$
hypercharge, with the other hypercharges set equal to zero (see Table~\ref{tab:Field_content}),
where $Y=Y_{1}+Y_{2}+Y_{3}$ is equal to SM weak hypercharge. Anomalies
cancel separately for each family, as in the SM, but without family
replication. The TH gauge group is broken down to the SM via appropriate
SM singlet scalars, which however carry family hypercharges. We denote
these fields linking the family hypercharges as \textit{hyperons}.
The TH group could be broken down to the SM in different ways, however
we motivate the following symmetry breaking pattern,\allowdisplaybreaks[0]
\begin{alignat}{1}
 & SU(3)_{c}\times SU(2)_{L}\times{\displaystyle U(1)_{Y_{1}}\times U(1)_{Y_{2}}\times U(1)_{Y_{3}}}\nonumber \\
 & {\displaystyle \overset{v_{12}}{\rightarrow}SU(3)_{c}\times SU(2)_{L}\times U(1)_{Y_{1}+Y_{2}}\times U(1)_{Y_{3}}}\label{eq:Symmetry_Breaking}\\
 & {\displaystyle \overset{v_{23}}{\rightarrow}SU(3)_{c}\times SU(2)_{L}\times U(1)_{Y_{1}+Y_{2}+Y_{3}}\,,}\nonumber 
\end{alignat}
\allowdisplaybreaks This choice is well supported by symmetry arguments
that will have phenomenological consequences: At high energies, the
TH group discriminates between the three SM fermion families, explicitly
breaking the approximate $U(3)^{5}$ flavour symmetry of the SM. At
a heavy scale $v_{12}$, the first and second hypercharges are broken
down to their diagonal subgroup, and the associated $Z'$ boson potentially
mediates dangerous 1-2 flavour-changing neutral currents (FCNCs).
Nevertheless, the gauge group below the scale $v_{12}$ preserves
an accidental $U(2)^{5}$ flavour symmetry. The groups $U(1)_{Y_{1}+Y_{2}}\times U(1)_{Y_{3}}$
are broken down to their diagonal subgroup (SM hypercharge) at a scale
$v_{23}$, and the associated $Z'$ boson is protected from mediating
the most dangerous FCNCs thanks to the $U(2)^{5}$ symmetry. In this
manner, the most dangerous FCNCs are suppressed by the heavier scale
$v_{12}$, while the scale $v_{23}$ can be very low with interesting
phenomenological implications. We will see that dynamics connected
to the scale $v_{12}$ will play a role in the origin of the family
hierarchy $m_{1}/m_{2}$, while dynamics connected to the scale $v_{23}$
will play a role in the origin of $m_{2}/m_{3}$. The distribution
of the various scales and the approximate flavour symmetries that
apply at each scale are summarised in the diagram of Fig.~\ref{fig:-Multiscale-picture-flavour}.

Provided that the SM Higgs only carries third family hypercharge,
$H(\mathbf{1},\mathbf{2})_{(0,0,-\frac{1}{2})}$, then the third family
Yukawa couplings are allowed at renormalisable level and an accidental
$U(2)^{5}$ flavour symmetry acting on the light families emerges
in the Yukawa sector, 
\begin{equation}
\mathcal{L}=y_{t}Q_{3}\tilde{H}u_{3}^{c}+y_{b}Q_{3}Hd_{3}^{c}+y_{\tau}L_{3}He_{3}^{c}+\mathrm{h.c.}
\end{equation}
where $\tilde{H}$ is the $CP$-conjugate of $H$. This setup already
provides an explanation for the smallness of light fermion masses
with respect to the third family, along with the smallness of quark
mixing, as they all must arise from non-renormalisable operators which
minimally break the $U(2)^{5}$ symmetry. Although this is a good
first order description of the SM spectrum, the question of why the
bottom and tau fermions are much lighter than the top remains unanswered,
and assuming only a single Higgs doublet, a tuning of order 2\% for
the bottom coupling and of 1\% for the tau coupling would be required.
Given that $m_{s,\mu}\propto\lambda^{5}m_{t}$ while $m_{c}\propto\lambda^{3}m_{t}$,
this setup also requires to generate a stronger fermion hierarchy
in the down and charged lepton sectors with respect to the up sector,
unless the tuning in the bottom and tau couplings is extended to the
second family. As we shall see shortly, the $U(1)_{Y}^{3}$ model
(and very likely a more general set of theories of flavour based on
the family decomposition of the SM group) predicts a similar mass
hierarchy for all charged sectors.

Due to the above considerations, it seems natural to consider a type
II two Higgs doublet model (2HDM), where both Higgs doublets only
carry third family hypercharge,
\begin{equation}
H_{u}(\mathbf{1},\mathbf{2})_{(0,0,\frac{1}{2})},\ \ \ \ H_{d}(\mathbf{1},\mathbf{2})_{(0,0,-\frac{1}{2})}\,,
\end{equation}
where as usual for a type II 2HDM, FCNCs can be forbidden by e.g.~a
$Z_{2}$ discrete symmetry or by supersymmetry (not necessarily low
scale), which we however do not specify in order to preserve the bottom-up
spirit of this work. In any case, $\tan\beta=v_{u}/v_{d}\sim\lambda^{-2}\approx20$,
which is compatible with current data (see e.g.~\cite{Beniwal:2022kyv,deGiorgi:2023wjh}),
will provide the hierarchy between the top and bottom/tau masses with
all dimensionless couplings being $\mathcal{O}(1)$. Such an overall
hierarchy between the down and charged leptons sector with respect
to the up sector is extended to all families, providing a better description
of second family charged fermion masses as we shall see.\allowdisplaybreaks

\section{Charged fermion masses and mixing\label{sec:Charged-fermion-masses-mixing}}

\subsection{Lessons from the spurion formalism\label{subsec:Lessons-from-the}}

In all generality, we introduce $U(2)^{5}$-breaking \textit{spurions}
$\Phi$ in the Yukawa matrices of charged fermions
\begin{flalign}
{\cal L} & =\begin{pmatrix}Q_{1} & Q_{2} & Q_{3}\end{pmatrix}\begin{pmatrix}{\Phi}(\frac{1}{2},0,-\frac{1}{2}) & {\Phi}(-\frac{1}{6},\frac{2}{3},-\frac{1}{2}) & {\Phi}(-\frac{1}{6},0,\frac{1}{6})\\
{\Phi}(\frac{2}{3},-\frac{1}{6},-\frac{1}{2}) & {\Phi}(0,\frac{1}{2},-\frac{1}{2}) & {\Phi}(0,-\frac{1}{6},\frac{1}{6})\\
{\Phi}(\frac{2}{3},0,-\frac{2}{3}) & {\Phi}(0,\frac{2}{3},-\frac{2}{3}) & 1
\end{pmatrix}\begin{pmatrix}u_{1}^{c}\\
u_{2}^{c}\\
u_{3}^{c}
\end{pmatrix}H_{u}\nonumber \\
 & +\begin{pmatrix}Q_{1} & Q_{2} & Q_{3}\end{pmatrix}\begin{pmatrix}{\Phi}(-\frac{1}{2},0,\frac{1}{2}) & {\Phi}(-\frac{1}{6},-\frac{1}{3},\frac{1}{2}) & {\Phi}(-\frac{1}{6},0,\frac{1}{6})\\
{\Phi}(-\frac{1}{3},-\frac{1}{6},\frac{1}{2}) & {\Phi}(0,-\frac{1}{2},\frac{1}{2}) & {\Phi}(0,-\frac{1}{6},\frac{1}{6})\\
{\Phi}(-\frac{1}{3},0,\frac{1}{3}) & {\Phi}(0,-\frac{1}{3},\frac{1}{3}) & 1
\end{pmatrix}\begin{pmatrix}d_{1}^{c}\\
d_{2}^{c}\\
d_{3}^{c}
\end{pmatrix}H_{d}\label{eq:Spurions}\\
 & +\begin{pmatrix}L_{1} & L_{2} & L_{3}\end{pmatrix}\begin{pmatrix}{\Phi}(-\frac{1}{2},0,\frac{1}{2}) & {\Phi}(\frac{1}{2},-1,\frac{1}{2}) & {\Phi}(\frac{1}{2},0,-\frac{1}{2})\\
{\Phi}(-1,\frac{1}{2},\frac{1}{2}) & {\Phi}(0,-\frac{1}{2},\frac{1}{2}) & {\Phi}(0,\frac{1}{2},-\frac{1}{2})\\
{\Phi}(-1,0,1) & {\Phi}(0,-1,1) & 1
\end{pmatrix}\begin{pmatrix}e_{1}^{c}\\
e_{2}^{c}\\
e_{3}^{c}
\end{pmatrix}H_{d}\,+\mathrm{h.c.}\,,\nonumber 
\end{flalign}
where each spurion carries non-trivial charge assignments under $U(1)_{Y}^{3}$.
In an effective field theory (EFT) approach, each spurion above can
be matched to specific ratios of hyperons $\phi_{i}$ over EFT cut-off
scales $\Lambda_{i}$, i.e. 
\begin{equation}
\Phi=\frac{\phi_{1}...\phi_{n}}{\Lambda_{1}...\Lambda_{n}}\,,
\end{equation}
where we have suppressed dimensionless couplings. The choice of hyperons
and $\Lambda_{i}$ above carries all the model dependence.

Assuming that the cut-off scales $\Lambda_{i}$ of the EFT are universal,
i.e.~that all $\Lambda_{i}$ are common to all charged sectors, then
the spurion formalism reveals some general considerations about the
origin of charged fermion masses and mixing: 
\begin{itemize}
\item The same spurions (up to conjugation) appear in the diagonal entries
of all matrices. Therefore, unless texture zeros are introduced in
specific models, this means that the masses of second family fermions
are likely to be degenerate up to dimensionless couplings, and the
same discussion applies to first family fermions. This motivates again
the addition of the second Higgs doublet or an alternative mechanism
in order to generate the hierarchy between the charm mass and the
lighter strange and muon masses. 
\item The same spurions appear in the (2,3) entries of the up and down Yukawa
matrices. Therefore, the 2-3 mixing in both the up and down sectors
is expected to be of a similar size, giving no predictions about the
alignment of the CKM element $V_{cb}$. The similar argument applies
to 1-3 mixing and $V_{ub}$. 
\item The spurions in the (1,2) entry of the up and down matrices are different.
Therefore, specific models have the potential to give predictions
about the alignment of the CKM element $V_{us}$. 
\item The same spurion (up to conjugation) that enters in all (2,2) entries
also populates the (2,3) entry of the charged lepton Yukawa matrix.
Similarly, the same spurion (up to conjugation) that enters in all
(1,1) entries also populates the (1,3) entry of the charged lepton
Yukawa matrix. In general, this predicts left-handed $\mu-\tau$ ($e-\tau$)
mixing of $\mathcal{O}(m_{2}/m_{3})$ ($\mathcal{O}(m_{1}/m_{3})$),
unless texture zeros are introduced in specific models (see Section~\ref{subsec:From-Spurions}).
This leads to a sizable enhancement of LFV $\tau\rightarrow\mu$ and
$\tau\rightarrow e$ transitions above the SM predictions, mediated
by heavy $Z'$ bosons in the model (see Section~\ref{sec:Phenomenology}). 
\item The spurions in the lower off-diagonal entries of the Yukawa matrices
all carry independent charge assignments, so right-handed fermion
mixing is model-dependent and can be different in all charged sectors. 
\end{itemize}
In the following, we go beyond the spurion formalism and introduce
different sets of hyperons. As we shall wee, the hyperons will provide
small $U(2)^{5}$-breaking effects via non-renormalisable operators,
leading to the masses of first and second family charged fermions,
along with CKM mixing. In the next few subsections, we will describe
example scenarios which provide a good description of charged fermion
masses and mixing.

\subsection{From spurions to hyperons \label{subsec:From-Spurions}}

The physical origin of the spurions of the previous subsection will
correspond to new Higgs scalar fields that break the $U(1)_{Y}^{3}$
symmetry, which we call \textit{hyperons}. The hyperons induce small
$U(2)^{5}$-breaking effects at the non-renormalisable level that
will lead to the masses and mixings of charged fermions. As the most
straightforward scenario, we could promote the spurions in the diagonal
entries of the matrices in Eq.~\eqref{eq:Spurions} to hyperons,
along with the off-diagonal spurions in the upper half of the down
matrix\footnote{Notice that the same spurions enter in both the (1,3) and (2,3) entries
of the up and down matrices in Eq.~\eqref{eq:Spurions}.}. In an EFT approach, the set of hyperons that we have assumed generates
the following Yukawa matrices, 
\begin{flalign}
{\cal L}^{d\leq5} & =\begin{pmatrix}Q_{1} & Q_{2} & Q_{3}\end{pmatrix}\begin{pmatrix}{\phi}_{\ell13}^{(\frac{1}{2},0,-\frac{1}{2})}/\Lambda & 0 & {\phi}_{q13}^{(-\frac{1}{6},0,\frac{1}{6})}/\Lambda\\
0 & {\phi}_{\ell23}^{(0,\frac{1}{2},-\frac{1}{2})}/\Lambda & {\phi}_{q23}^{(0,-\frac{1}{6},\frac{1}{6})}/\Lambda\\
0 & 0 & 1
\end{pmatrix}\begin{pmatrix}u_{1}^{c}\\
u_{2}^{c}\\
u_{3}^{c}
\end{pmatrix}H_{u}\\
 & +\begin{pmatrix}Q_{1} & Q_{2} & Q_{3}\end{pmatrix}\begin{pmatrix}{\tilde{\phi}}_{\ell13}^{(-\frac{1}{2},0,\frac{1}{2})}/\Lambda & {\phi}_{d12}^{(-\frac{1}{6},-\frac{1}{3},\frac{1}{2})}/\Lambda & {\phi}_{q13}^{(-\frac{1}{6},0,\frac{1}{6})}/\Lambda\\
0 & {\tilde{\phi}}_{\ell23}^{(0,-\frac{1}{2},\frac{1}{2})}/\Lambda & {\phi}_{q23}^{(0,-\frac{1}{6},\frac{1}{6})}/\Lambda\\
0 & 0 & 1
\end{pmatrix}\begin{pmatrix}d_{1}^{c}\\
d_{2}^{c}\\
d_{3}^{c}
\end{pmatrix}H_{d}\\
 & +\begin{pmatrix}L_{1} & L_{2} & L_{3}\end{pmatrix}\begin{pmatrix}{\tilde{\phi}}_{\ell13}^{(-\frac{1}{2},0,\frac{1}{2})}/\Lambda & 0 & {\phi}_{\ell13}^{(\frac{1}{2},0,-\frac{1}{2})}/\Lambda\\
0 & {\tilde{\phi}}_{\ell23}^{(0,-\frac{1}{2},\frac{1}{2})}/\Lambda & {\phi}_{\ell23}^{(0,\frac{1}{2},-\frac{1}{2})}/\Lambda\\
0 & 0 & 1
\end{pmatrix}\begin{pmatrix}e_{1}^{c}\\
e_{2}^{c}\\
e_{3}^{c}
\end{pmatrix}H_{d}\,+\mathrm{h.c.}\,,
\end{flalign}
where the universal scale $\Lambda$ is the high cut-off scale of
the EFT, and we ignore the $\mathcal{O}(1)$ dimensionless couplings
of each entry. Although we have chosen only the specific set of hyperons
shown, leaving some zeros in the matrices, these zeros may be filled
in by higher order operators with dimension larger than 5, which so
far we are ignoring.

When the hyperons develop VEVs, assumed to be smaller than the cut-off
scale $\Lambda$, then each entry of the matrix will receive a suppressed
numerical effective coupling given by ratios of the form $\left\langle {\phi}\right\rangle /\Lambda$,
whose values can be assumed arbitrarily. Having the freedom to choose
arbitrary VEVs for each hyperon, the Yukawa matrices above could provide
a good first order description of charged fermion masses and CKM mixing.
We choose to fix the $\left\langle {\phi}\right\rangle /\Lambda$
ratios in terms of powers of the Wolfenstein parameter $\lambda\simeq0.224$,
obtaining 
\begin{flalign}
{\cal L} & =\begin{pmatrix}u_{1} & u_{2} & u_{3}\end{pmatrix}\begin{pmatrix}\lambda^{6} & 0 & \lambda^{3}\\
0 & \lambda^{3} & \lambda^{2}\\
0 & 0 & 1
\end{pmatrix}\begin{pmatrix}u_{1}^{c}\\
u_{2}^{c}\\
u_{3}^{c}
\end{pmatrix}\frac{v_{\mathrm{SM}}}{\sqrt{2}}\label{eq:Texture_a}\\
 & +\begin{pmatrix}d_{1} & d_{2} & d_{3}\end{pmatrix}\begin{pmatrix}\lambda^{6} & \lambda^{4} & \lambda^{3}\\
0 & \lambda^{3} & \lambda^{2}\\
0 & 0 & 1
\end{pmatrix}\begin{pmatrix}d_{1}^{c}\\
d_{2}^{c}\\
d_{3}^{c}
\end{pmatrix}\lambda^{2}\frac{v_{\mathrm{SM}}}{\sqrt{2}}\label{eq:Texture_b}\\
 & +\begin{pmatrix}e_{1} & e_{2} & e_{3}\end{pmatrix}\begin{pmatrix}\lambda^{6} & 0 & \lambda^{6}\\
0 & \lambda^{3} & \lambda^{3}\\
0 & 0 & 1
\end{pmatrix}\begin{pmatrix}e_{1}^{c}\\
e_{2}^{c}\\
e_{3}^{c}
\end{pmatrix}\lambda^{2}\frac{v_{\mathrm{SM}}}{\sqrt{2}}\,+\mathrm{h.c.}\label{eq:Texture_c}
\end{flalign}
As anticipated from the spurion formalism, the alignment of $V_{cb}$
and $V_{ub}$ is not predicted by the model. In contrast, the model
predicts a relevant left-handed $\mu-\tau$ ($e-\tau$) mixing connected
to the same hyperon that provides the second family (first family)
effective Yukawa couplings. Thanks to the addition of the second Higgs
doublet, the model successfully explains third and second family fermion
masses with $\mathcal{O}(1)$ dimensionless couplings. The down-quark
and electron masses are also reasonably explained, although the up-quark
mass is naively a factor $\mathcal{O}(\lambda^{-1.5})$ larger than
current data. Notice that so far we are only assuming one universal
cut-off scale $\Lambda$, while in realistic models several cut-off
scales $\Lambda$ may be associated to different messengers in the
UV theory, which could provide a larger suppression for the up-quark
effective coupling. Therefore, within the limitations of our EFT approach,
the description of charged fermion masses given by the set of Eqs.~(\ref{eq:Texture_a}-\ref{eq:Texture_c})
is very successful.

As an alternative example, one could also consider a model where the
(1,1) spurion in Eq.~\eqref{eq:Spurions} is not promoted to hyperon,
but instead all the spurions in the (1,2) and (2,1) entries are promoted,
so that the Yukawa matrices show an exact texture zero in the (1,1)
entry,
\begin{flalign}
{\cal L}^{d\leq5} & =\begin{pmatrix}Q_{1} & Q_{2} & Q_{3}\end{pmatrix}\begin{pmatrix}0 & {\phi}_{u12}^{(-\frac{1}{6},\frac{2}{3},-\frac{1}{2})} & {\phi}_{q13}^{(-\frac{1}{6},0,\frac{1}{6})}\\
{\phi}_{u21}^{(\frac{2}{3},-\frac{1}{6},-\frac{1}{2})} & {\phi}_{\ell23}^{(0,\frac{1}{2},-\frac{1}{2})} & {\phi}_{q23}^{(0,-\frac{1}{6},\frac{1}{6})}\\
0 & 0 & 1
\end{pmatrix}\begin{pmatrix}u_{1}^{c}\\
u_{2}^{c}\\
u_{3}^{c}
\end{pmatrix}H_{u}\\
 & +\begin{pmatrix}Q_{1} & Q_{2} & Q_{3}\end{pmatrix}\begin{pmatrix}0 & {\phi}_{d12}^{(-\frac{1}{6},-\frac{1}{3},\frac{1}{2})} & {\phi}_{q13}^{(-\frac{1}{6},0,\frac{1}{6})}\\
{\phi}_{d21}^{(-\frac{1}{3},-\frac{1}{6},\frac{1}{2})} & {\tilde{\phi}}_{\ell23}^{(0,-\frac{1}{2},\frac{1}{2})} & {\phi}_{q23}^{(0,-\frac{1}{6},\frac{1}{6})}\\
0 & 0 & 1
\end{pmatrix}\begin{pmatrix}d_{1}^{c}\\
d_{2}^{c}\\
d_{3}^{c}
\end{pmatrix}H_{d}\\
 & +\begin{pmatrix}L_{1} & L_{2} & L_{3}\end{pmatrix}\begin{pmatrix}0 & {\phi}_{e12}^{(\frac{1}{2},-1,\frac{1}{2})} & 0\\
{\phi}_{e21}^{(-1,\frac{1}{2},\frac{1}{2})} & {\tilde{\phi}}_{\ell23}^{(0,-\frac{1}{2},\frac{1}{2})} & {\phi}_{\ell23}^{(0,\frac{1}{2},-\frac{1}{2})}\\
0 & 0 & 1
\end{pmatrix}\begin{pmatrix}e_{1}^{c}\\
e_{2}^{c}\\
e_{3}^{c}
\end{pmatrix}H_{d}\,+\mathrm{h.c.}\,,
\end{flalign}
where we have omitted the high cut-off $\Lambda$ suppressing each
dimension-5 operator above. The VEV over $\Lambda$ ratios of the
new hyperons can be fixed by the requirement of addressing first family
fermion masses, obtaining Yukawa matrices with texture zeros given
by
\begin{flalign}
{\cal L} & =\begin{pmatrix}u_{1} & u_{2} & u_{3}\end{pmatrix}\begin{pmatrix}0 & \lambda^{5} & \lambda^{3}\\
\lambda^{5.5} & \lambda^{3} & \lambda^{2}\\
0 & 0 & 1
\end{pmatrix}\begin{pmatrix}u_{1}^{c}\\
u_{2}^{c}\\
u_{3}^{c}
\end{pmatrix}\frac{v_{\mathrm{SM}}}{\sqrt{2}}\\
 & +\begin{pmatrix}d_{1} & d_{2} & d_{3}\end{pmatrix}\begin{pmatrix}0 & \lambda^{4} & \lambda^{3}\\
\lambda^{4} & \lambda^{3} & \lambda^{2}\\
0 & 0 & 1
\end{pmatrix}\begin{pmatrix}d_{1}^{c}\\
d_{2}^{c}\\
d_{3}^{c}
\end{pmatrix}\lambda^{2}\frac{v_{\mathrm{SM}}}{\sqrt{2}}\\
 & +\begin{pmatrix}e_{1} & e_{2} & e_{3}\end{pmatrix}\begin{pmatrix}0 & \lambda^{5} & 0\\
\lambda^{4.4} & \lambda^{3} & \lambda^{3}\\
0 & 0 & 1
\end{pmatrix}\begin{pmatrix}e_{1}^{c}\\
e_{2}^{c}\\
e_{3}^{c}
\end{pmatrix}\lambda^{2}\frac{v_{\mathrm{SM}}}{\sqrt{2}}\,+\mathrm{h.c.}\,,
\end{flalign}
which provide an even better description of first family fermion masses
than the original simplified model. Notice that in this scenario,
a sizable left-handed $e-\tau$ mixing is no longer predicted.

We conclude that the most straightforward choices of hyperons, motivated
by the spurion formalism, already provide a good description of charged
fermion masses and mixings. However, these simplified models leave
some questions unanswered. Given that we are assuming the symmetry
breaking highlighted in Eq.~\eqref{eq:Symmetry_Breaking} and Fig.~\ref{fig:-Multiscale-picture-flavour},
we notice that there are no hyperons breaking the first and second
hypercharges down to their diagonal subgroup, and we would expect
those to play a role in the origin of fermion hierarchies and mixing.
Moreover, in the simplified models introduced so far, several hyperons
display unexplained large hierarchies of VEVs whose values are assumed
\textit{a posteriori} to fit the fermion masses. Given that all these
hyperons participate in the 23-breaking step of Eq.~\eqref{eq:Symmetry_Breaking},
we would expect all of them to develop VEVs at a similar scale, rather
than the hierarchical scales assumed. This motivates further model
building. In the following subsections we discuss a couple of example
models which address these issues.

\subsection{Model 1: Minimal case with three hyperons\label{subsec:Model-1:-Minimal}}

We introduce here the following set of three hyperons,
\begin{equation}
{\phi}_{\ell23}^{(0,\frac{1}{2},-\frac{1}{2})}\,,\qquad{\phi}_{q23}^{(0,-\frac{1}{6},\frac{1}{6})}\,,\qquad{\phi}_{q12}^{(-\frac{1}{6},\frac{1}{6},0)}\,.
\end{equation}
Following the EFT approach of the previous subsection, we now analyse
the effective Yukawa matrices obtained by combining the SM charged
fermions, the Higgs doublets and the hyperons, in a tower of non-renormalisable
operators preserving the $U(1)_{Y}^{3}$ gauge symmetry, 
\begin{flalign}
{\cal L} & =\begin{pmatrix}Q_{1} & Q_{2} & Q_{3}\end{pmatrix}\begin{pmatrix}\tilde{\phi}_{q12}^{3}{\phi}_{\ell23} & {\phi}_{q12}{\phi}_{\ell23} & {\phi}_{q12}{\phi}_{q23}\\
\tilde{\phi}_{q12}^{4}{\phi}_{\ell23} & {\phi}_{\ell23} & {\phi}_{q23}\\
\tilde{\phi}_{q12}^{4}{\phi}_{\ell23}\tilde{\phi}_{q23} & {\phi}_{\ell23}\tilde{\phi}_{q23} & 1
\end{pmatrix}\begin{pmatrix}u_{1}^{c}\\
u_{2}^{c}\\
u_{3}^{c}
\end{pmatrix}H_{u}\\
 & +\begin{pmatrix}Q_{1} & Q_{2} & Q_{3}\end{pmatrix}\begin{pmatrix}\phi_{q12}^{3}\tilde{\phi}_{\ell23} & {\phi}_{q12}\tilde{\phi}_{\ell23} & {\phi}_{q12}{\phi}_{q23}\\
{\phi}_{q12}^{2}\tilde{\phi}_{\ell23} & \tilde{\phi}_{\ell23} & {\phi}_{q23}\\
{\phi}_{q12}^{2}{\phi}_{q23}^{2} & {\phi}_{q23}^{2} & 1
\end{pmatrix}\begin{pmatrix}d_{1}^{c}\\
d_{2}^{c}\\
d_{3}^{c}
\end{pmatrix}H_{d}\\
 & +\begin{pmatrix}L_{1} & L_{2} & L_{3}\end{pmatrix}\begin{pmatrix}\phi_{q12}^{3}\tilde{\phi}_{\ell23} & \tilde{\phi}_{q12}^{3}\tilde{\phi}_{\ell23} & \tilde{\phi}_{q12}^{3}{\phi}_{\ell23}\\
\phi_{q12}^{6}\tilde{\phi}_{\ell23} & \tilde{\phi}_{\ell23} & {\phi}_{\ell23}\\
\phi_{q12}^{6}\tilde{\phi}_{\ell23}^{2} & \tilde{\phi}_{\ell23}^{2} & 1
\end{pmatrix}\begin{pmatrix}e_{1}^{c}\\
e_{2}^{c}\\
e_{3}^{c}
\end{pmatrix}H_{d}\,+\mathrm{h.c.}\,,
\end{flalign}
where the powers of $\Lambda$ in the denominator and the dimensionless
couplings of each entry are not shown. Once the hyperons above develop
VEVs, we obtain very economical and efficient Yukawa textures for
modeling the observed pattern of SM Yukawa couplings. In particular,
the masses of second family fermions arise at dimension 5 in the EFT,
while first family masses have an extra suppression as they arise
from dimension-8 operators. Regarding CKM mixing, 2-3 quark mixing
leading to $V_{cb}$ arises from dimension-5 operators, while $V_{ub}$
has an extra mild suppression as it arises from dimension-6 operators.
In all cases, right-handed fermion mixing is suppressed with respect
to left-handed mixing. This is a highly desirable feature, given the
strong phenomenological constraints on right-handed flavour-changing
currents \cite{UTfit:2007eik,Isidori:2014rba}, which may be mediated
by heavy $Z'$ bosons arising from the symmetry breaking of $U(1)_{Y}^{3}$
(see Section~\ref{sec:Phenomenology}).

In good approximation, quark mixing leading to $V_{us}$ arises as
the ratio of the (1,2) and (2,2) entries of the quark matrices above,
therefore we expect
\begin{equation}
\frac{\left\langle {\phi}_{q12}\right\rangle }{\Lambda}\sim V_{us}\simeq\lambda\,,
\end{equation}
where $\lambda=\sin\theta_{C}\simeq0.224$. In a similar manner, we
can fix the ratio $\left\langle {\phi}_{q23}\right\rangle /\Lambda$
by reproducing the observed $V_{cb}$
\begin{equation}
\frac{\left\langle {\phi}_{q23}\right\rangle }{\Lambda}\sim V_{cb}\simeq\lambda^{2}\,.
\end{equation}
Given that both $\left\langle {\phi}_{q23}\right\rangle $ and $\left\langle {\phi}_{\ell23}\right\rangle $
play a role in the last step of the symmetry breaking cascade (see
Fig.~\ref{fig:-Multiscale-picture-flavour}), it is expected that
both VEVs live at a similar scale, although they are not expected
to be degenerate but differ by an $\mathcal{O}(1)$ factor. This way,
we are free to choose
\begin{equation}
\frac{\left\langle {\phi}_{\ell23}\right\rangle }{\Lambda}\sim\frac{m_{c}}{m_{t}}\sim\lambda^{3}\,,
\end{equation}
which, given that $\left\langle H_{d}\right\rangle $ provides an
extra suppression of $\mathcal{O}(\lambda^{2})$ for down-quarks and
charged lepton Yukawas, allows to predict all second family masses
with $\mathcal{O}(1)$ dimensionless couplings. In contrast with the
simplified models of Section~\ref{subsec:From-Spurions}, this model
provides all the 23-breaking VEVs at the same scale, plus a larger
12-breaking VEV, following a mild hierarchy given by $v_{23}/v_{12}\sim\lambda$.
This way, the symmetry breaking of the $U(1)_{Y}^{3}$ gauge group
proceeds just like in Eq.~\eqref{eq:Symmetry_Breaking} and Fig.~\ref{fig:-Multiscale-picture-flavour},
as desired.

Having fixed all the hyperon VEVs with respect to $\Lambda$, now
we are able to write the full mass matrices for each sector in terms
of the Wolfenstein parameter $\lambda$, 
\begin{flalign}
{\cal L} & =\begin{pmatrix}u_{1} & u_{2} & u_{3}\end{pmatrix}\begin{pmatrix}\lambda^{6} & \lambda^{4} & \lambda^{3}\\
\lambda^{7} & \lambda^{3} & \lambda^{2}\\
\lambda^{9} & \lambda^{5} & 1
\end{pmatrix}\begin{pmatrix}u_{1}^{c}\\
u_{2}^{c}\\
u_{3}^{c}
\end{pmatrix}\frac{v_{\mathrm{SM}}}{\sqrt{2}}\\
 & +\begin{pmatrix}d_{1} & d_{2} & d_{3}\end{pmatrix}\begin{pmatrix}\lambda^{6} & \lambda^{4} & \lambda^{3}\\
\lambda^{5} & \lambda^{3} & \lambda^{2}\\
\lambda^{6} & \lambda^{4} & 1
\end{pmatrix}\begin{pmatrix}d_{1}^{c}\\
d_{2}^{c}\\
d_{3}^{c}
\end{pmatrix}\lambda^{2}\frac{v_{\mathrm{SM}}}{\sqrt{2}}\\
 & +\begin{pmatrix}e_{1} & e_{2} & e_{3}\end{pmatrix}\begin{pmatrix}\lambda^{6} & \lambda^{6} & \lambda^{6}\\
\lambda^{9} & \lambda^{3} & \lambda^{3}\\
\lambda^{12} & \lambda^{6} & 1
\end{pmatrix}\begin{pmatrix}e_{1}^{c}\\
e_{2}^{c}\\
e_{3}^{c}
\end{pmatrix}\lambda^{2}\frac{v_{\mathrm{SM}}}{\sqrt{2}}\,+\mathrm{h.c.}\,.
\end{flalign}
We can see that this setup provides a reasonable description of charged
fermion masses and mixing. Although the up and down-quark masses are
slightly off by $\mathcal{O}(\lambda)$ factors, we remember that
we are only assuming one universal cut-off scale $\Lambda$, while
in realistic models several scales $\Lambda$ may be associated to
different messengers in the UV theory, further improving the fit of
first family quark masses, as discussed in Section~\ref{subsec:From-Spurions}.
All things considered, the description of fermion masses seems very
efficient, considering the limitations of our EFT framework.

However, the model does not predict the alignment of $V_{us}$. Moreover,
we also notice that right-handed $s-d$ mixing is just mildly suppressed
as $s_{12}^{d_{R}}\simeq\mathcal{O}(\lambda^{2})$ in this model.
Given the stringent bounds over left-right scalar operators contributing
to $K^{0}-\bar{K}^{0}$ meson mixing \cite{UTfit:2007eik,Isidori:2014rba}
(which might arise in this kind of models as we shall see in Section~\ref{sec:Phenomenology}),
the scale $v_{12}$ can be pushed far above the TeV if $V_{us}$ originates
from the down sector. From a phenomenological point of view, it would
be interesting to find models which give clear predictions about the
alignment of $V_{us}$, and ideally provide a more efficient suppression
of right-handed quark mixing. We shall see in the next subsection
that this can be achieved by minimally extending the set of hyperons
of this model.

\subsection{Model 2: Five hyperons for a more predictive setup\label{subsec:Model-2:-Five}}

Model 1 proposed in the previous subsection, despite its simplicity
and minimality, does not give clear predictions about the alignment
of the CKM matrix. Heavy $Z'$ bosons arising from the symmetry breaking
of $U(1)_{Y}^{3}$ have the potential to mediate contributions to
$K^{0}-\bar{K}^{0}$ meson mixing, which could set a lower bound over
the scale of $U(1)_{Y}^{3}$-breaking, but such contributions depend
on the alignment of $V_{us}$. Moreover, the largest contributions
to $K^{0}-\bar{K}^{0}$ mixing depend both on the alignment of $V_{us}$
and on right-handed $s-d$ mixing, which is just mildly suppressed
in Model 1. Therefore, we propose here a similar model with slightly
extended hyperon content that can account for a clear prediction about
the alignment of $V_{us}$, plus a more efficient suppression of right-handed
fermion mixing. We consider here the hyperons 
\begin{equation}
{\phi}_{\ell23}^{(0,\frac{1}{2},-\frac{1}{2})},\qquad{\phi}_{q23}^{(0,-\frac{1}{6},\frac{1}{6})},\qquad{\phi}_{q13}^{(-\frac{1}{6},0,\frac{1}{6})},\qquad{\phi}_{d12}^{(-\frac{1}{6},-\frac{1}{3},\frac{1}{2})},\qquad{\phi}_{e12}^{(\frac{1}{4},-\frac{1}{4},0)}.
\end{equation}
With this set of hyperons, the effective Yukawa couplings in the EFT
are (suppressing as usual powers of $\Lambda$ and dimensionless couplings)
\begin{flalign}
{\cal L} & =\begin{pmatrix}Q_{1} & Q_{2} & Q_{3}\end{pmatrix}\begin{pmatrix}\phi_{e12}^{2}\tilde{\phi}_{\ell23} & {\phi}_{q13}{\tilde{\phi}}_{q23}{\phi}_{\ell23} & {\phi}_{q13}\\
{\phi}_{e12}^{2}{\tilde{\phi}}_{q13}{\phi}_{q23}{\phi}_{\ell23} & {\phi}_{\ell23} & {\phi}_{q23}\\
{\phi}_{e12}^{2}{\tilde{\phi}}_{q13}{\phi}_{\ell23} & {\phi}_{\ell23}\tilde{\phi}_{q23} & 1
\end{pmatrix}\begin{pmatrix}u_{1}^{c}\\
u_{2}^{c}\\
u_{3}^{c}
\end{pmatrix}H_{u}\\
 & +\begin{pmatrix}Q_{1} & Q_{2} & Q_{3}\end{pmatrix}\begin{pmatrix}\phi_{e12}^{2}\tilde{\phi}_{\ell23} & {\phi}_{d12} & {\phi}_{q13}\\
{\phi}_{q13}^{2}{\phi}_{q23} & \tilde{\phi}_{\ell23} & {\phi}_{q23}\\
{\phi}_{q13}^{2} & {\phi}_{q23}^{2} & 1
\end{pmatrix}\begin{pmatrix}d_{1}^{c}\\
d_{2}^{c}\\
d_{3}^{c}
\end{pmatrix}H_{d}\\
 & +\begin{pmatrix}L_{1} & L_{2} & L_{3}\end{pmatrix}\begin{pmatrix}\tilde{\phi}_{e12}^{2}\tilde{\phi}_{\ell23} & \phi_{e12}^{2}\tilde{\phi}_{\ell23} & \phi_{e12}^{2}{\phi}_{\ell23}\\
\tilde{\phi}_{e12}^{4}\tilde{\phi}_{\ell23} & \tilde{\phi}_{\ell23} & {\phi}_{\ell23}\\
\tilde{\phi}_{e12}^{4}\tilde{\phi}_{\ell23}^{2} & {\tilde{\phi}}_{\ell23}^{2} & 1
\end{pmatrix}\begin{pmatrix}e_{1}^{c}\\
e_{2}^{c}\\
e_{3}^{c}
\end{pmatrix}H_{d}\,+\mathrm{h.c.}\,.
\end{flalign}
Following the same approach as with Model 1, we assign the following
powers of $\lambda$ to the VEV over $\Lambda$ ratios in order to
reproduce fermion masses and CKM mixing, 
\begin{equation}
\frac{\left\langle {\phi}_{\ell23}\right\rangle }{\Lambda}=\frac{\left\langle {\phi}_{q13}\right\rangle }{\Lambda}\simeq\lambda^{3}\,,\quad\frac{\left\langle {\phi}_{q23}\right\rangle }{\Lambda}\simeq\lambda^{2}\,,\quad\frac{\left\langle {\phi}_{d12}\right\rangle }{\Lambda}\simeq\lambda^{4}\,,\quad\frac{\left\langle {\phi}_{e12}\right\rangle }{\Lambda}\simeq\lambda\,.
\end{equation}
Although it would seem that in this scenario there exists a mild hierarchy
between 23-breaking VEVs of $\mathcal{O}(\lambda^{2})$, since ${\phi}_{d12}$
only appears in the 12 entry of the down matrix, it would be very
reasonable that the dimensionless coupling in that entry provides
a factor $\lambda$ suppression, such that all 23-breaking VEVs live
at the same scale. The largest VEV is still the 12-breaking one, which
is now associated to the hyperon ${\phi}_{e12}$, and the mild hierarchy
between scales remains as $v_{12}/v_{23}\simeq\lambda$. With these
assignments of VEVs over $\Lambda$ ratios, the Yukawa textures are
given by 
\begin{flalign}
{\cal L} & =\begin{pmatrix}u_{1} & u_{2} & u_{3}\end{pmatrix}\begin{pmatrix}\lambda^{5} & \lambda^{8} & \lambda^{3}\\
\lambda^{10} & \lambda^{3} & \lambda^{2}\\
\lambda^{7} & \lambda^{5} & 1
\end{pmatrix}\begin{pmatrix}u_{1}^{c}\\
u_{2}^{c}\\
u_{3}^{c}
\end{pmatrix}\frac{v_{\mathrm{SM}}}{\sqrt{2}}\\
 & +\begin{pmatrix}d_{1} & d_{2} & d_{3}\end{pmatrix}\begin{pmatrix}\lambda^{5} & \lambda^{4} & \lambda^{3}\\
\lambda^{8} & \lambda^{3} & \lambda^{2}\\
\lambda^{6} & \lambda^{4} & 1
\end{pmatrix}\begin{pmatrix}d_{1}^{c}\\
d_{2}^{c}\\
d_{3}^{c}
\end{pmatrix}\lambda^{2}\frac{v_{\mathrm{SM}}}{\sqrt{2}}\\
 & +\begin{pmatrix}e_{1} & e_{2} & e_{3}\end{pmatrix}\begin{pmatrix}\lambda^{5} & \lambda^{5} & \lambda^{5}\\
\lambda^{7} & \lambda^{3} & \lambda^{3}\\
\lambda^{10} & \lambda^{6} & 1
\end{pmatrix}\begin{pmatrix}e_{1}^{c}\\
e_{2}^{c}\\
e_{3}^{c}
\end{pmatrix}\lambda^{2}\frac{v_{\mathrm{SM}}}{\sqrt{2}}\,+\mathrm{h.c.}\,.
\end{flalign}
Just like in Model 1, this model provides a compelling description
of all charged fermion masses and mixing. Notice that this scenario
provides a very efficient suppression of right-handed fermion mixing.
Moreover, it is clear that here $V_{us}$ mixing originates from the
down sector, providing a more predictive setup than Model 1, which
will be useful for phenomenological purposes as discussed in Section~\ref{sec:Phenomenology}.

Finally we comment that the Higgs doublets and hyperons can mediate
FCNCs such as $K^{0}-\bar{K}^{0}$ mixing. In the present model, this
may arise from tree-level exchange of $\phi_{d12}$ hyperons which
can mediate down-strange transitions. Such a coupling originates from
$Q_{1}\phi_{d12}d_{2}^{c}H_{d}/\Lambda$ which will lead to a suppressed
down-strange coupling of order $\lambda^{2}v_{\mathrm{SM}}/\Lambda\approx10^{-5}$
(taking $\Lambda\approx100\;\mathrm{TeV}$ as expected if $v_{23}\approx\mathcal{O}(\mathrm{TeV})$).
Assuming the mass of the hyperons to be at the scale $v_{23}$, then
we expect these FCNCs to be under control. Since we assume a type
II 2HDM, tree-level Higgs doublets exchange contributions are forbidden,
and FCNCs mediated by the Higgs doublets can only proceed via their
mixing with hyperons, carrying therefore an extra suppression via
the mixing angle, along with the suppression of hyperon couplings
already discussed. In more general models of this kind, such contributions
to $K^{0}-\bar{K}^{0}$ mixing could be even further suppressed depending
on the order of the operator and the alignment of $V_{us}$, and in
all generality we expect all hyperon couplings to be suppressed by
at least a factor $v_{\mathrm{SM}}/\Lambda\approx10^{-3}$. A more
detailed study of the phenomenology of the scalar sector in this general
class of models is beyond the scope of this work.

\section{Neutrino masses and mixing\label{sec:Neutrino-masses-and-Mixing}}

\subsection{General considerations and spurion formalism}

The origin of neutrino masses and mixing requires a dedicated analysis
due to their particular properties. We start by introducing $U(2)^{5}$-breaking
spurions (carrying inverse of mass dimension) for the Weinberg operator
\begin{equation}
\mathcal{L}_{\mathrm{Weinberg}}=\begin{pmatrix}L_{1} & L_{2} & L_{3}\end{pmatrix}\begin{pmatrix}{\Phi}(1,0,-1) & {\Phi}(\frac{1}{2},\frac{1}{2},-1) & {\Phi}(\frac{1}{2},0,-\frac{1}{2})\\
{\Phi}(\frac{1}{2},\frac{1}{2},-1) & {\Phi}(0,1,-1) & {\Phi}(0,\frac{1}{2},-\frac{1}{2})\\
{\Phi}(\frac{1}{2},0,-\frac{1}{2}) & {\Phi}(0,\frac{1}{2},-\frac{1}{2}) & 1
\end{pmatrix}\begin{pmatrix}L_{1}\\
L_{2}\\
L_{3}
\end{pmatrix}H_{u}H_{u}\,,
\end{equation}
which reveals that, as expected, the $U(2)^{5}$ approximate symmetry
is naively present in the neutrino sector as well. As a consequence,
one generally expects one neutrino to be much heavier than the others,
displaying tiny mixing with the other neutrino flavours. In the spirit
of the type I seesaw mechanism, one could think of adding a $U(1)_{Y}^{3}$
singlet neutrino as $N(0,0,0)$. Such a singlet neutrino can only
couple to the third family active neutrino at renormalisable level,
i.e.~$\mathcal{L}_{N}\supset L_{3}H_{u}N+m_{N}NN$, where all fermion
fields are written in a left-handed 2-component convention. This way,
the coupling $L_{2}H_{u}N$, which is required for large atmospheric
neutrino mixing, can only arise at the non-renormalisable level. Therefore,
it is expected to be suppressed with respect to $L_{3}H_{u}N$. This
seems to be inconsistent with large atmospheric neutrino mixing, at
least within the validity of our EFT framework. As anticipated before,
this is a consequence of the accidental $U(2)^{5}$ flavour symmetry
delivered by the TH model.

Given such general considerations, we conclude that in order to obtain
neutrino masses and mixing from the type I seesaw mechanism, it is
required to add SM singlet neutrinos that carry tri-hypercharges (but
whose hypercharges add up to zero). These neutrinos will allow to
introduce $U(2)^{5}$-breaking operators similar for all neutrino
flavours, providing a mechanism to obtain the adequate neutrino mixing
in a natural way. In order to cancel gauge anomalies, the most simple
option is that these singlet neutrinos are vector-like. As a consequence,
they can obtain their mass from the unspecified vector-like mass terms
and from the VEVs of hyperons in the specific model.

Remarkably, if a given SM singlet neutrino $N_{\mathrm{atm}}$ carrying
non-trivial hypercharges provides the couplings $\mathcal{L}_{N_{\mathrm{atm}}}\supset L_{3}H_{u}N_{\mathrm{atm}}+L_{2}H_{u}N_{\mathrm{atm}}$,
as required to explain atmospheric mixing, then the conjugate neutrino
will always couple to $L_{3}$ as $L_{3}H_{u}\overline{N}_{\mathrm{atm}}$,
but not necessarily to $L_{2}$. As shown in Appendix~\ref{sec:General-formalism-for-seesaw},
when the terms $L_{3}H_{u}\overline{N}_{\mathrm{atm}}$ enter the
seesaw mechanism, they lead to a hierarchical effective neutrino mass
matrix proportional to the vector-like mass terms, while the terms
involving SM singlet neutrinos like $N_{\mathrm{atm}}$ lead to a
neutrino mass matrix where all entries are $\mathcal{O}(1)$ and proportional
to the VEVs of the hyperons. Therefore, the simplest way to explain
the observed pattern of large neutrino mixing requires that the vector-like
masses are of similar or smaller order than the VEVs of the hyperons,
leading to a \textit{low scale} seesaw mechanism if the VEVs of the
hyperons are not very large (which is consistent with current data
as discussed in Section~\ref{sec:Phenomenology}).

\subsection{Example of successful neutrino mixing from the seesaw mechanism\label{subsec:Example-of-seesaw}}

In the following, we provide an example scenario which reproduces
the observed pattern of neutrino mixing, as a proof of principle.
According to the discussion in the previous subsection, in order to
implement a type I seesaw mechanism that delivers large neutrino mixing,
we need to add vector-like neutrinos that carry tri-hypercharges (but
whose hypercharges add up to zero). We also need to introduce hyperons
that will provide small Dirac mass terms for the active neutrinos
in the form of non-renormalisable operators. Under these considerations,
we start by adding the following vector-like neutrino\footnote{We remind the reader that in our convention, all fermion fields including
$N_{\mathrm{atm}}$ and $\overline{N}_{\mathrm{atm}}$ are left-handed. } and hyperon 
\begin{equation}
N_{\mathrm{atm}}^{(0,\frac{1}{4},-\frac{1}{4})}\,,\qquad\overline{N}_{\mathrm{atm}}^{(0,-\frac{1}{4},\frac{1}{4})}\,,\qquad\phi_{\mathrm{atm}}^{(0,\frac{1}{4},-\frac{1}{4})}\,,
\end{equation}
where the charge assignments are chosen to provide large \textit{atmospheric
neutrino mixing}. This way, we can write the following non-renormalisable
operators along with the Majorana and vector-like masses of $N_{\mathrm{atm}}$,
\begin{align}
\mathcal{L}_{N_{\mathrm{atm}}}\supset & \frac{1}{\Lambda_{\mathrm{atm}}}(\phi_{\mathrm{atm}}L_{2}+\tilde{\phi}_{\mathrm{atm}}L_{3})H_{u}N_{\mathrm{atm}}+\frac{\phi_{\mathrm{atm}}}{\Lambda_{\mathrm{atm}}}L_{3}H_{u}\overline{N}_{\mathrm{atm}}\label{eq:Latm}\\
 & +\phi_{\ell23}N_{\mathrm{atm}}N_{\mathrm{atm}}+\tilde{\phi}_{\ell23}\overline{N}_{\mathrm{atm}}\overline{N}_{\mathrm{atm}}+M_{N_{\mathrm{atm}}}\overline{N}_{\mathrm{atm}}N_{\mathrm{atm}}\,,\nonumber 
\end{align}
where we have ignored the $\mathcal{O}(1)$ dimensionless couplings,
and the hyperon $\phi_{\ell23}^{(0,\frac{1}{2},-\frac{1}{2})}$ is
already present in both Model 1 and Model 2 for the charged fermion
sector. In a similar spirit, we introduce another vector-like neutrino
and other hyperons in order to obtain large \textit{solar neutrino
mixing} 
\begin{equation}
N_{\mathrm{sol}}^{(\frac{1}{4},\frac{1}{4},-\frac{1}{2})}\,,\qquad\overline{N}_{\mathrm{sol}}^{(-\frac{1}{4},-\frac{1}{4},\frac{1}{2})}\,,\qquad\phi_{\mathrm{sol}}^{(-\frac{1}{2},-\frac{1}{2},1)},\qquad\phi_{\nu13}^{(-\frac{1}{4},-\frac{1}{4},\frac{1}{2})}\,,
\end{equation}
which provide the following non-renormalisable operators and mass
terms, 
\begin{align}
\mathcal{L}_{N_{\mathrm{sol}}}\supset & \frac{1}{\Lambda_{\mathrm{sol}}}(\phi_{e12}L_{1}+\tilde{\phi}_{e12}L_{2}+\phi_{\nu13}L_{3})H_{u}N_{\mathrm{sol}}+\frac{\phi_{\nu13}}{\Lambda_{\mathrm{sol}}}L_{3}H_{u}\overline{N}_{\mathrm{sol}}\label{eq:Lsol}\\
 & +\phi_{\mathrm{sol}}N_{\mathrm{sol}}N_{\mathrm{sol}}+\tilde{\phi}_{\mathrm{sol}}\overline{N}_{\mathrm{sol}}\overline{N}_{\mathrm{sol}}+M_{N_{\mathrm{sol}}}\overline{N}_{\mathrm{sol}}N_{\mathrm{sol}}\,,\nonumber 
\end{align}
where we have ignored again the $\mathcal{O}(1)$ dimensionless couplings,
and the hyperon $\phi_{e12}^{(\frac{1}{2},-\frac{1}{2},0)}$ is already
present in Model 2 for the charged fermion sector. The hyperon $\phi_{\nu13}$
(which will eventually populate the (1,3) entry of the effective neutrino
mass matrix) is not required to obtain non-zero reactor mixing, which
would already arise from the other operators, but it is required in
order to have enough free parameters to fit all observed neutrino
mixing angles and mass splittings.

Notice that the vector-like neutrinos $N_{\mathrm{atm}}$ and $N_{\mathrm{sol}}$
get contributions to their masses from the VEVs of the hyperons $\phi_{\ell23}$
and $\phi_{\mathrm{sol}}$, respectively, which we denote generically
as $v_{23}$ since they both take part in the 23-breaking step of
Eq.~\eqref{eq:Symmetry_Breaking}. In addition, $N_{\mathrm{atm}}$
and $N_{\mathrm{sol}}$ get contributions to their masses from the
unspecified vector-like mass terms, that we generically denote as
$M_{\mathrm{VL}}$. As shown in Appendix~\ref{sec:General-formalism-for-seesaw}
(see Eq.~\eqref{eq:Seesaw_General}), given that the conjugate neutrinos
$\overline{N}_{\mathrm{atm}}$ and $\overline{N}_{\mathrm{sol}}$
only couple to the third family, the seesaw formula reveals that the
effective neutrino mass matrix $m_{\nu}$ receives two main contributions: 
\begin{itemize}
\item A contribution proportional to $v_{23}$ which populates all entries
of $m_{\nu}$ with $\mathcal{O}(1)$ terms. 
\item A contribution proportional to the vector-like masses $M_{\mathrm{VL}}$,
which populates only the third row and column entries of $m_{\nu}$
with $\mathcal{O}(1)$ terms and the others are zero. 
\end{itemize}
Therefore, if $M_{\mathrm{VL}}\gg v_{23}$, then after the seesaw
mechanism the light effective Majorana neutrino mass matrix $m_{\nu}$
will have only the third row and column being non-zero (to good approximation),
which is inconsistent with the observed pattern of neutrino mixing
and mass splittings. Instead, if $M_{\mathrm{VL}}\apprle v_{23}$,
then the contribution proportional $v_{23}$ dominates the seesaw
mechanism. In this case, the conjugate neutrinos $\overline{N}_{\mathrm{atm}}$
and $\overline{N}_{\mathrm{sol}}$ become irrelevant for the seesaw
mechanism, and $m_{\nu}$ can be obtained by considering only the
presence of the SM singlet neutrinos $N_{\mathrm{atm}}$ and $N_{\mathrm{sol}}$
and applying the seesaw formula. We construct the Dirac and Majorana
matrices (ignoring $\mathcal{O}(1)$ dimensionless couplings) as
\begin{equation}
m_{D}=\left(
\global\long\def\arraystretch{0.7}%
\begin{array}{@{}llc@{}}
 & \multicolumn{1}{c@{}}{\phantom{\!\,}N_{\mathrm{sol}}} & \phantom{\!\,}N_{\mathrm{atm}}\\
\cmidrule(l){2-3}\left.L_{1}\right| & \frac{\phi_{e12}}{\Lambda_{\mathrm{sol}}} & 0\\
\left.L_{2}\right| & \frac{\tilde{\phi}_{e12}}{\Lambda_{\mathrm{sol}}} & \frac{\phi_{\mathrm{atm}}}{\Lambda_{\mathrm{atm}}}\\
\left.L_{3}\right| & \frac{\phi_{\nu13}}{\Lambda_{\mathrm{sol}}} & \frac{\tilde{\phi}_{\mathrm{atm}}}{\Lambda_{\mathrm{atm}}}
\end{array}\right)H_{u}\,,\qquad M_{N}=\left(
\global\long\def\arraystretch{0.7}%
\begin{array}{@{}llc@{}}
 & \multicolumn{1}{c@{}}{\phantom{\!\,}N_{\mathrm{sol}}} & \phantom{\!\,}N_{\mathrm{atm}}\\
\cmidrule(l){2-3}\left.\;\,N_{\mathrm{sol}}\right| & \phi_{\mathrm{sol}} & 0\\
\left.N_{\mathrm{atm}}\right| & 0 & \phi_{\ell23}
\end{array}\right)\,.\label{seesaw}
\end{equation}

We could have included the $U(1)_{Y}^{3}$ singlet neutrino $N(0,0,0)$,
but as discussed in the previous subsection, its contributions to
the Weinberg operator may be suppressed by its large Majorana mass
$m_{N}$, resulting in possibly negligible contributions to the seesaw
mechanism. We are therefore free to assume that such a neutrino $N(0,0,0)$,
if exists, is in any case decoupled from the seesaw, while the atmospheric
and solar SM singlet neutrinos $N_{\mathrm{atm}}$ and $N_{\mathrm{sol}}$
could yield dominant and subdominant contributions, resulting in a
natural normal neutrino mass hierarchy as in sequential dominance
\cite{King:1998jw,King:1999mb,King:2002nf}.

More generally, after applying the seesaw formula using Eq.~\eqref{seesaw},
we obtain the light effective Majorana neutrino mass matrix (ignoring
again $\mathcal{O}(1)$ dimensionless couplings) as 
\begin{flalign}
m_{\nu} & \simeq m_{D}M_{N}^{-1}m_{D}^{\text{\ensuremath{\mathrm{T}}}}\\
 & =\left(\begin{array}{ccc}
\frac{1}{\Lambda_{\mathrm{sol}}^{2}}\frac{\phi_{e12}^{2}}{\phi_{\mathrm{sol}}} & \frac{1}{\Lambda_{\mathrm{sol}}^{2}}\frac{\phi_{e12}\tilde{\phi}_{e12}}{\phi_{\mathrm{sol}}} & \frac{1}{\Lambda_{\mathrm{sol}}^{2}}\frac{\phi_{e12}\phi_{\nu13}}{\phi_{\mathrm{sol}}}\\
\frac{1}{\Lambda_{\mathrm{sol}}^{2}}\frac{\phi_{e12}\tilde{\phi}_{e12}}{\phi_{\mathrm{sol}}} & \frac{1}{\Lambda_{\mathrm{atm}}^{2}}\frac{\phi_{\mathrm{atm}}^{2}}{\phi_{\ell23}}+\frac{1}{\Lambda_{\mathrm{sol}}^{2}}\frac{\tilde{\phi}_{e12}^{2}}{\phi_{\mathrm{sol}}} & \frac{1}{\Lambda_{\mathrm{atm}}^{2}}\frac{\phi_{\mathrm{atm}}\tilde{\phi}_{\mathrm{atm}}}{\phi_{\ell23}}+\frac{1}{\Lambda_{\mathrm{sol}}^{2}}\frac{\tilde{\phi}_{e12}\phi_{\nu13}}{\phi_{\mathrm{sol}}}\\
\frac{1}{\Lambda_{\mathrm{sol}}^{2}}\frac{\phi_{e12}\phi_{\nu13}}{\phi_{\mathrm{sol}}} & \frac{1}{\Lambda_{\mathrm{atm}}^{2}}\frac{\phi_{\mathrm{atm}}\tilde{\phi}_{\mathrm{atm}}}{\phi_{\ell23}}+\frac{1}{\Lambda_{\mathrm{sol}}^{2}}\frac{\tilde{\phi}_{e12}\phi_{\nu13}}{\phi_{\mathrm{sol}}} & \frac{1}{\Lambda_{\mathrm{atm}}^{2}}\frac{\phi_{\mathrm{atm}}^{2}}{\phi_{\ell23}}+\frac{1}{\Lambda_{\mathrm{sol}}^{2}}\frac{\phi_{\nu13}^{2}}{\phi_{\mathrm{sol}}}
\end{array}\right)H_{u}H_{u}\,.\nonumber 
\end{flalign}
Given the symmetry breaking pattern of the model shown in Eq.~\eqref{eq:Symmetry_Breaking},
we take $\left\langle {\phi}_{e12}\right\rangle \simeq\mathcal{O}(v_{12})$
and $\left\langle \phi_{\nu13}\right\rangle \approx\left\langle \phi_{\ell23}\right\rangle \approx\left\langle \phi_{\mathrm{atm}}\right\rangle \approx\left\langle \phi_{\mathrm{sol}}\right\rangle \approx\mathcal{O}(v_{23})$\footnote{In the calculations that follow, we assume for simplicity that these
VEVs are equal. However, the same conclusions hold as long as the
VEVs vary by $\mathcal{O}(1)$ factors, which is the natural expectation.}. Motivated by our discussion of the charged fermion sector (see Section~\ref{sec:Charged-fermion-masses-mixing}),
we consider the relation $v_{23}/v_{12}\simeq\lambda$. By inserting
such VEVs, we obtain
\begin{equation}
m_{\nu}\simeq\left(\begin{array}{ccc}
0 & 0 & 0\\
0 & 1 & 1\\
0 & 1 & 1
\end{array}\right)v_{23}\frac{H_{u}H_{u}}{\Lambda_{\mathrm{atm}}^{2}}\,+\left(\begin{array}{ccc}
1 & 1 & \lambda\\
1 & 1 & \lambda\\
\lambda & \lambda & \lambda^{2}
\end{array}\right)v_{23}\frac{H_{u}H_{u}}{\lambda^{2}\Lambda_{\mathrm{sol}}^{2}}\,.
\end{equation}
If $\Lambda_{\mathrm{sol}}=\Lambda_{\mathrm{atm}}$, we observe that
there exists a mild hierarchy of order $\lambda^{2}$ between the
12 and 23 sectors in the matrix above. Considering the dimensionless
coefficients that we have ignored so far, the numerical diagonalisation
of $m_{\nu}$ would require some parameters of $\mathcal{O}(0.01)$
in order to explain the observed neutrino mixing angles and mass splittings
\cite{deSalas:2020pgw,Gonzalez-Garcia:2021dve}. The situation can
be improved if we assume a mild hierarchy between cut-off scales $\Lambda_{\mathrm{atm}}/\Lambda_{\mathrm{sol}}\simeq\lambda$,
obtaining to leading order for each entry (ignoring dimensionless
coefficients), \clearpage
\begin{equation}
m_{\nu}\simeq\left(\begin{array}{ccc}
1 & 1 & \lambda\\
1 & 1 & 1\\
\lambda & 1 & 1
\end{array}\right)v_{23}\frac{v_{\mathrm{SM}}^{2}}{\Lambda_{\mathrm{atm}}^{2}}\,,
\end{equation}
where we have introduced the SM VEV as $\left\langle H_{u}\right\rangle =v_{\mathrm{SM}}$
(ignoring the factor $1/\sqrt{2}$). Considering now the dimensionless
coefficients in the matrix above, we find that numerical diagonalisation
can accommodate all the observed neutrino mixing angles and mass splittings
\cite{deSalas:2020pgw,Gonzalez-Garcia:2021dve} with $\mathcal{O}(1)$
parameters, and we are able to reproduce both normal and inverted
ordered scenarios.

Notice that we have been driven to a scenario where the vector-like
neutrinos get Majorana masses from the VEVs of hyperons in the model.
Furthermore, the vector-like masses necessarily have to be of the
same or smaller order than the VEVs of the hyperons in order to explain
the observed pattern of neutrino mixing. Therefore, in the particular
example included in this section, the vector-like neutrinos get a
mass at the scale $v_{23}$ of the 23-breaking step in Fig.~\eqref{fig:-Multiscale-picture-flavour},
which could happen at a relatively low scale as we shall see in Section~\ref{sec:Phenomenology}.
As a consequence, the vector-like neutrinos involved in the seesaw
mechanism are expected to be relatively light, and the high energy
cut-offs of the EFT $\Lambda_{\mathrm{atm}}$ and $\Lambda_{\mathrm{sol}}$
are expected to provide most the suppression for tiny neutrino masses.
As anticipated before, due to the $U(2)^{5}$ flavour symmetry provided
by the TH model, we have been driven to a low scale seesaw in order
to predict the observed pattern of neutrino mixing.

\section{Phenomenology\label{sec:Phenomenology}}

\subsection{Couplings of the heavy $Z'$ bosons to fermions \label{subsec:Couplings_Zprime}}

In Sections~\ref{sec:Charged-fermion-masses-mixing} and \ref{sec:Neutrino-masses-and-Mixing}
we have discussed examples of $U(1)_{Y}^{3}$ models which provide
a compelling description of all fermion masses and mixings, and we
have highlighted model-independent features which are intrinsic to
the $U(1)_{Y}^{3}$ framework. Under well-motivated arguments, we
have assumed that the symmetry breaking pattern of the $U(1)_{Y}^{3}$
group down to the SM is described by Eq.~\eqref{eq:Symmetry_Breaking}
and Fig.~\ref{fig:-Multiscale-picture-flavour}, in such a way that
at a high scale $v_{12}$, the group $U(1)_{Y_{1}}\times U(1)_{Y_{2}}$
is broken down to its diagonal subgroup. The remaining group $U(1)_{Y_{1}+Y_{2}}\times U(1)_{Y_{3}}$
is broken down to SM hypercharge at a lower scale $v_{23}$. The hierarchy
between the scales $v_{12}$ and $v_{23}$ generally plays a role
on the origin of flavour hierarchies in the SM, although in specific
models we have found that a mild hierarchy $v_{23}/v_{12}\simeq\lambda$
is enough.

A massive gauge boson $Z'_{12}$ is predicted to live at the higher
scale $v_{12}$, displaying \textit{intrinsically} flavour non-universal
couplings to the first two families of SM fermions. Similarly, another
massive boson $Z'_{23}$ lives at the lower scale $v_{23}$. The pattern
of symmetry breaking is such that $Z'_{23}$ has flavour universal
couplings to first and second family fermions, while the couplings
to the third family are intrinsically different. In the following,
we include the coupling matrices in family space (from the covariant
derivatives in Appendices~\ref{sec:High-scale-symmetry} and~\ref{sec:Low-scale-symmetry}),
ignoring fermion mass mixing,
\begin{flalign}
 & \mathcal{L}_{Z'_{12}}\supset Y_{\psi_{L,R}}\overline{\psi}_{L,R}\gamma^{\mu}\left(\begin{array}{ccc}
-g_{1}\sin\theta_{12} & 0 & 0\\
0 & g_{2}\cos\theta_{12} & 0\\
0 & 0 & 0
\end{array}\right)\psi_{L,R}Z'_{12\mu}\,, & \hspace{-1em}\hspace{-1em}\sin\theta_{12}=\frac{g_{1}}{\sqrt{g_{1}^{2}+g_{2}^{2}}}\,,\label{eq:Z12_couplings}
\end{flalign}
\begin{flalign}
 & \mathcal{L}_{Z'_{23}}\supset Y_{\psi_{L,R}}\overline{\psi}_{L,R}\gamma^{\mu}\left(\begin{array}{ccc}
-g_{12}\sin\theta_{23} & 0 & 0\\
0 & -g_{12}\sin\theta_{23} & 0\\
0 & 0 & g_{3}\cos\theta_{23}
\end{array}\right)\psi_{L,R}Z'_{23\mu}\,, & \hspace{-1em}\hspace{-1em}\sin\theta_{23}=\frac{g_{12}}{\sqrt{g_{12}^{2}+g_{3}^{2}}}\,,\label{eq:Z23_couplings}
\end{flalign}
where $Y_{\psi_{L,R}}$ is the SM hypercharge of $\psi_{L,R}$\footnote{Note that we have departed from our 2-component and purely left-handed
convention, used in the rest of the paper, to use instead a 4-component
left-right convention, which is more familiar in phenomenological
studies.}, where $\psi$ is a 3-component column vector containing the three
families $\psi=u^{i},d^{i},e^{i},\nu^{i}$. Explicitly, $\psi_{L}=u_{L}^{i},d_{L}^{i},e_{L}^{i},\nu_{L}^{i}$
with $Y_{\psi_{L}}=1/6,1/6,-1/2,-1/2$, and $\psi_{R}=u_{R}^{i},d_{R}^{i},e_{R}^{i}$
with $Y_{\psi_{R}}=2/3,-1/3,-1$, respectively, ignoring couplings
to the SM singlet neutrinos discussed in the previous section\footnote{Note that such low scale SM singlet neutrinos may be observable via
their gauge couplings to $Z'_{23}$, which can be obtained from the
covariant derivative in Eq.~\eqref{eq:cov_23}.}.

Including fermion mass mixing, we would have $\psi_{L,R}=V_{\psi_{L,R}}\hat{\psi}_{L,R}$
with $\hat{\psi}_{L,R}$ containing the mass eigenstates and $V_{\psi_{L,R}}$
being the \textit{non-generic} mixing matrices obtained after diagonalising
the Yukawa matrices for a given model. Notice that the couplings to\textit{
right-handed} fermions are larger since their hypercharges are generally
larger in magnitude than those of left-handed fermions. The SM hypercharge
gauge coupling $g_{Y}(M_{Z})\simeq0.36$ is entangled to the $g_{i}$
couplings via the matching conditions 
\begin{equation}
g_{Y}=\frac{g_{12}g_{3}}{\sqrt{g_{12}^{2}+g_{3}^{2}}}\,,\qquad\qquad\qquad\qquad g_{12}=\frac{g_{1}g_{2}}{\sqrt{g_{1}^{2}+g_{2}^{2}}}\,.\label{eq:gauge_couplings}
\end{equation}
The expressions above reveal a lower bound on the gauge couplings
$g_{i}\apprge g_{Y}$. Moreover, we may use the matching condition
with SM hypercharge to exchange $g_{12}$ in favour of $g_{3}$ and
$g_{Y}$ for the couplings of $Z'_{23}$,
\begin{equation}
\mathcal{L}_{Z'_{23}}\supset Y_{\psi_{L,R}}\overline{\psi}_{L,R}\gamma^{\mu}\left(\begin{array}{ccc}
-{\displaystyle \frac{g_{Y}^{2}}{\sqrt{g_{3}^{2}-g_{Y}^{2}}}} & 0 & 0\\
0 & -{\displaystyle \frac{g_{Y}^{2}}{\sqrt{g_{3}^{2}-g_{Y}^{2}}}} & 0\\
0 & 0 & \sqrt{g_{3}^{2}-g_{Y}^{2}}
\end{array}\right)\psi_{L,R}Z'_{23\mu}\,,
\end{equation}
Therefore, the phenomenology of $Z'_{23}$ can be completely described
in terms of its mass and the $g_{3}$ coupling. When $g_{3}$ is large,
$Z'_{23}$ is mostly coupled to the third family, while for $g_{3}$
small $Z'_{23}$ is mostly coupled to the first and second families.
In contrast, the couplings of $Z'_{12}$ need to be described in general
by two free gauge couplings. Here we choose $g_{1}$ and $g_{2}$,
but one may exchange one of these by e.g.~$g_{3}$ and $g_{Y}$ through
the matching conditions.

Throughout this work we have considered a bottom-up approach where
the $U(1)_{Y}^{3}$ model is just the next step in our understanding
of Nature, which reveals information about the origin of flavour,
but nevertheless is an EFT remnant of a more fundamental UV-complete
theory. In this spirit, we have studied the RGE evolution of the gauge
couplings $g_{i}$, obtaining that for $g_{i}(\mathrm{TeV})\simeq1$
the model can be extrapolated to the Planck scale (and beyond). Instead,
for $g_{i}(\mathrm{TeV})\simeq2$, a Landau pole is found at a scale
$\mathcal{O}(10^{4}\:\mathrm{TeV})$, which anyway seems like a reasonable
scale for an UV embedding, given that we expect the cut-off scale
of the effective Yukawa operators (see Section~\ref{sec:Charged-fermion-masses-mixing})
to be around $\mathcal{O}(10^{2}\:\mathrm{TeV})$ in order to provide
the required suppression for charged fermion masses. Therefore, in
order to protect the perturbativity of the model, we avoid considering
$g_{i}>2$ in the phenomenological analysis. Nevertheless, we highlight
a natural scenario where the three gauge couplings have a similar
size $g_{1}\simeq g_{2}\simeq g_{3}\simeq\sqrt{3}g_{Y}$, which could
be connected to a possible \textit{gauge unification}. This benchmark
is depicted as a dashed horizontal line in Figs.~\ref{fig:Zp12_plot}
and \ref{fig:Zp23_plot}.

\subsection{The high scale boson $Z'_{12}$}

In any implementation of the $U(1)_{Y}^{3}$ model, $Z'_{12}$ is
expected to mediate sizable tree-level transitions between first and
second generation left-handed quarks, either in the up or down sector
depending on the alignment of the CKM matrix predicted by the specific
model. Furthermore, our analysis in Section~\ref{sec:Charged-fermion-masses-mixing}
reveals that $U(1)_{Y}^{3}$ models generally predict non-vanishing
charged lepton mixing and mixing among right-handed quarks. This way,
contributions to $K^{0}-\bar{K}^{0}$ and $D^{0}-\bar{D}^{0}$ meson
mixing \cite{UTfit:2007eik,Isidori:2014rba}, along with CLFV processes
such as $\mu\rightarrow e\gamma$ \cite{PDG:2022ynf}, have the potential
to push the scale $v_{12}$ far above the $\mathrm{TeV}$. 
\begin{figure}[t]
\begin{centering}
\includegraphics[scale=0.5]{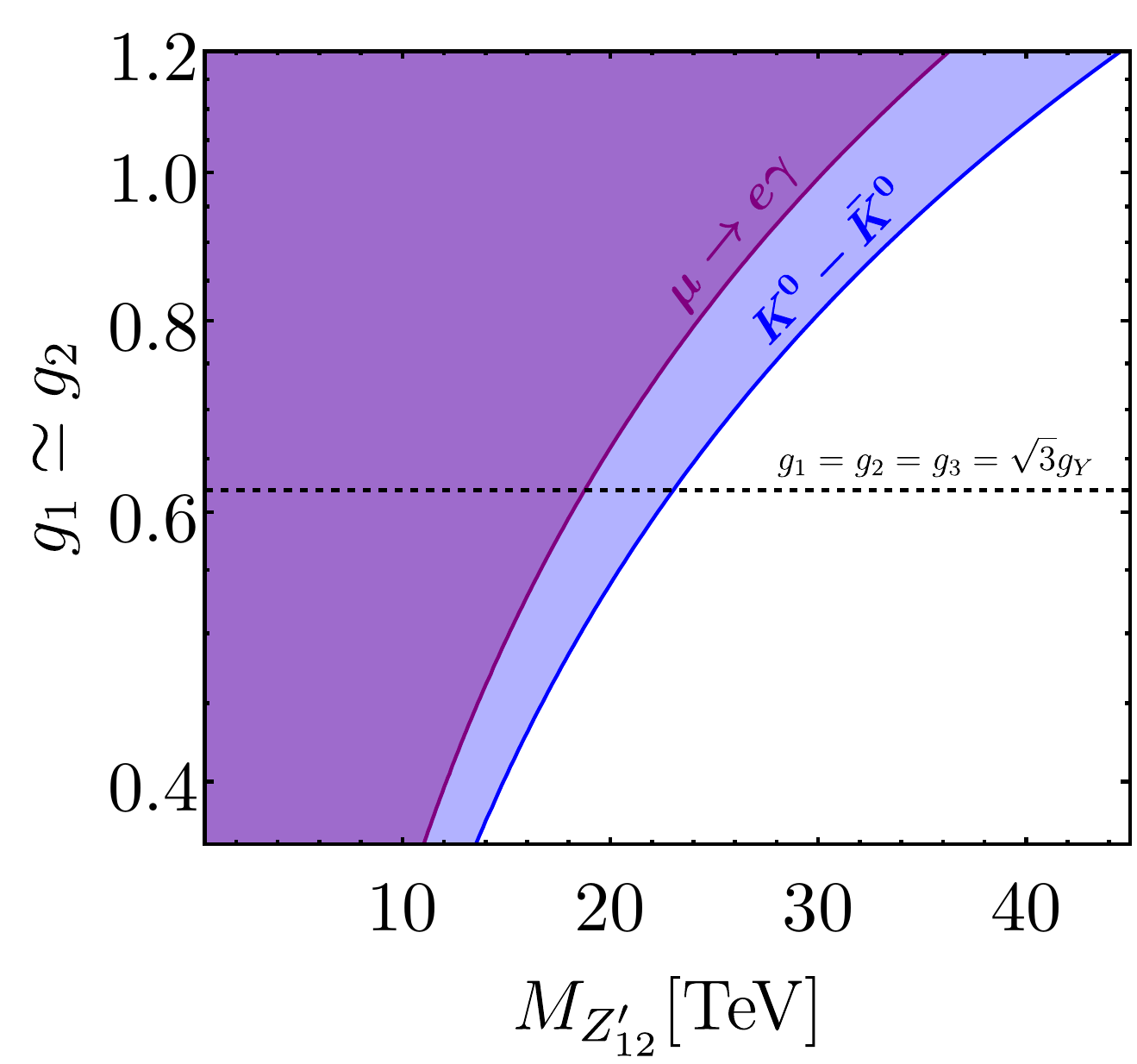} 
\par\end{centering}
\caption{Parameter space of the high scale breaking, where $M_{Z'_{12}}$ is
the mass of the heavy $Z'_{12}$ gauge boson and $g_{1}$, $g_{2}$
are the gauge couplings of the $U(1)_{Y_{1}}$ and $U(1)_{Y_{2}}$
groups, respectively. For simplicity, we assume $g_{1}$ and $g_{2}$
to be similar, and the non-generic fermion mixing predicted by Model
2 in Section~\ref{subsec:Model-2:-Five}. Shaded regions in the plot
depict 95\% CL exclusions over the parameter space. The dashed line
represents the natural benchmark $g_{1}\simeq g_{2}\simeq g_{3}\simeq\sqrt{3}g_{Y}$
motivated in the main text. \label{fig:Zp12_plot}}
\end{figure}

Being more specific, for Model 1 described in Section~\ref{subsec:Model-1:-Minimal}
we find the stringent bounds over $v_{12}$ to come from the scalar
and coloured operator $(\bar{s}_{L}^{\alpha}d_{R}^{\beta})(\bar{s}_{R}^{\beta}d_{L}^{\alpha})$
obtained after integrating out $Z'_{12}$ at tree-level (and applying
a Fierz rearrangement), which contributes to $K^{0}-\bar{K}^{0}$
mixing. Model 1 predicts the up and down left-handed mixings to be
similar up to dimensionless couplings, which must therefore play some
role in the alignment of the CKM matrix. In either case, mildly suppressed
right-handed $s-d$ mixing $s_{12}^{d_{R}}\simeq\mathcal{O}(\lambda^{2})$
is predicted. If $V_{us}$ originates mostly from the down sector,
then $K^{0}-\bar{K}^{0}$ mixing imposes the stringent bound $M_{Z'_{12}}>170\:\mathrm{TeV}$
for gauge couplings of $\mathcal{O}(0.5)$. Instead, if the dimensionless
coupling provides a mild suppression of $\mathcal{O}(0.1)$ in left-handed
$s-d$ mixing, such that $V_{us}$ originates mostly from the up sector,
then the bound is relaxed to $M_{Z'_{12}}>55\:\mathrm{TeV}$. We find
bounds from $D^{0}-\bar{D}^{0}$ mixing to be always weaker, even
if $V_{us}$ originates from the up sector, since right-handed up
mixing is strongly suppressed in Model 1.

In contrast with Model 1, Model 2 described in Section~\ref{subsec:Model-2:-Five}
provides a more predictive scenario where $V_{us}$ originates unambiguously
from the down sector. Here right-handed quark mixing is more suppressed,
obtaining $s_{12}^{d_{R}}\simeq\mathcal{O}(\lambda^{5})$. Nevertheless,
$K^{0}-\bar{K}^{0}$ mixing still imposes the strongest bounds over
the parameter case, as can be seen in Fig.~\ref{fig:Zp12_plot}.
In this case, the lower bound over the mass of $Z'_{12}$ can be as
low as 10-50 TeV, depending on the values of the gauge couplings.
We find the CLFV process $\mu\rightarrow e\gamma$ to provide a slightly
weaker bound over the parameter space, because charged lepton mixing
is generally suppressed with respect to quark mixing in Model 2. We
find the bound from $\mu\rightarrow3e$ to be very similar to the
bound from $\mu\rightarrow e\gamma$.

\subsection{The low scale boson $Z'_{23}$}

Given that the high scale symmetry breaking can be as low as 10-20
TeV for specific models, and considering the hierarchy of scales $v_{23}/v_{12}\simeq\lambda$
suggested by these specific models in Section~\ref{sec:Charged-fermion-masses-mixing},
it is possible to find the low scale breaking $v_{23}$ near the TeV.
Since $Z'_{23}$ features flavour universal couplings to the first
and second families, the stringent bounds from $K^{0}-\bar{K}^{0}$
mixing and $\mu\rightarrow e\gamma$ are avoided, in the spirit of
the \textit{GIM mechanism}. This way, $Z'_{23}$ can live at the TeV
scale, within the reach of the LHC and future colliders.

Any implementation of the $U(1)_{Y}^{3}$ model predicts small \textit{mixing}
between $Z'_{23}$ and the SM $Z$ boson given by the mixing angle
(see Appendix~\ref{sec:Low-scale-symmetry})\footnote{In this section we only discuss the impact of $Z-Z'_{23}$ gauge mixing,
while kinetic mixing is found to be negligible as long as the kinetic
mixing parameter is smaller than $\mathcal{O}(1)$ (see Appendix~\ref{sec:Low-scale-symmetry}).}
\begin{equation}
\mathrm{sin}\theta_{Z-Z'_{23}}=\frac{\sqrt{g_{3}^{2}-g_{Y}^{2}}}{\sqrt{g_{Y}^{2}+g_{L}^{2}}}\left(\frac{M_{Z}^{0}}{M_{Z'_{23}}^{0}}\right)^{2}\,,
\end{equation}
where $M_{Z}^{0}$ and $M_{Z'_{23}}^{0}$ are the masses of the $Z$
and $Z'_{23}$ bosons in the absence of mixing, respectively, and
$g_{L}$ is the gauge coupling of $SU(2)_{L}$. This mixing leads
to a \textit{small shift} on the mass of the $Z$ boson, which has
an impact on the so-called $\rho$ parameter 
\begin{equation}
\rho=\frac{M_{W}^{2}}{M_{Z}^{2}\cos^{2}\theta_{W}}=\frac{1}{1-(g_{3}^{2}-g_{Y}^{2})\left(\frac{v_{\mathrm{SM}}}{2M_{Z'_{23}}^{0}}\right)^{2}}\,,
\end{equation}
which is predicted as $\rho=1$ at tree-level in the SM. This is a
consequence of custodial symmetry in the Higgs potential, which is
explicitly broken in our model via $Z-Z'_{23}$ mixing leading to
deviations in $\rho$. The fact that in our model $M_{Z}$ is always
shifted to smaller values leads to $\rho>1$ at tree-level. Given
that $M_{Z}$ is commonly an input experimental parameter of the SM
used in the determination of $g_{Y}$ and $g_{L}$, the downward shift
of $M_{Z}$ with respect to the SM prediction would be seen from the
experimental point of view as an upward shift of $M_{W}$ with respect
to the SM prediction. Nevertheless, the experimental picture of $M_{W}$
is puzzling after the recent measurement by CDF \cite{CDF:2022hxs}.
This measurement points towards $M_{W}$ being larger than the SM
prediction with high significance, but it is in tension with the combination
of measurements by LHC, LEP and Tevatron D0 \cite{PDG:2022ynf}. Neglecting
the recent CDF measurement for the moment, current data\footnote{The current world average (without the latest CDF measurement) of
$M_{W}$ does not consider the very recent $M_{W}$ update by ATLAS~\cite{ATLAS:2023fsi}.
Given that the central value and the uncertainty of this measurement
are just slightly reduced with respect to the 2017 measurement \cite{ATLAS:2017rzl},
we do not expect a big impact over the world average.} provides $\rho=1.0003\pm0.0005$ \cite{PDG:2022ynf} (assuming that
both the oblique parameters $T$ and $S$ are non-zero, as we expect
in our model). We obtain the approximate bound $g_{3}/M_{Z'}<3.1\,\mathrm{TeV}$
at 95\% CL, which translates to an approximate bound over the mixing
angle of $\mathrm{sin}\theta_{Z-Z'_{23}}<0.001$. 
\begin{figure}[t]
\begin{centering}
\includegraphics[scale=0.5]{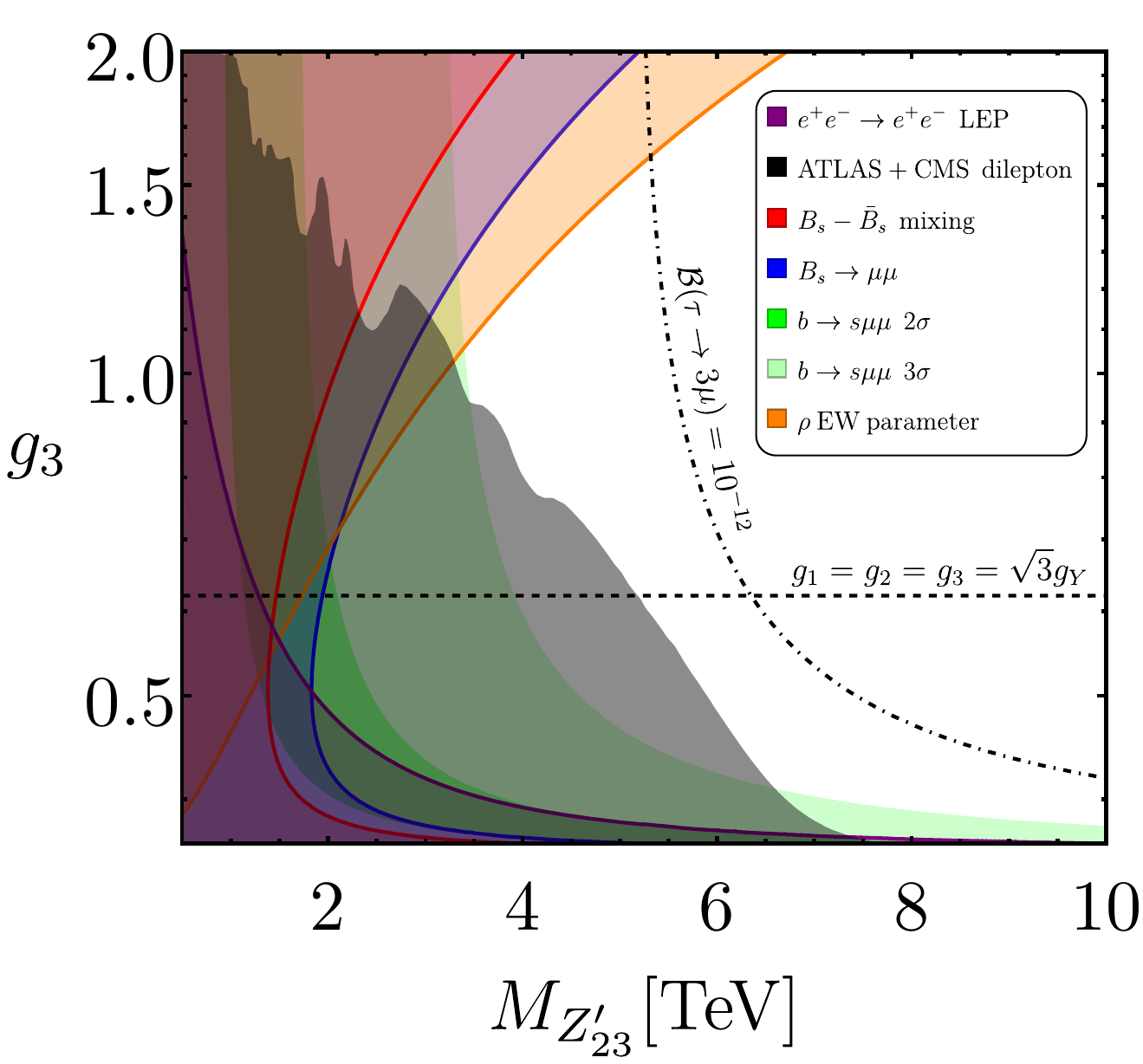} 
\par\end{centering}
\caption{Parameter space of the low scale breaking, where $M_{Z'_{23}}$ is
the mass of the heavy $Z'_{23}$ gauge boson and $g_{3}$ is the gauge
coupling of the $U(1)_{Y_{3}}$ group. The gauge coupling $g_{12}$
is fixed in terms of $g_{3}$ and $g_{Y}$ via Eq.~\eqref{eq:gauge_couplings},
and we consider the non-generic fermion mixing predicted by Model
2 in Section~\ref{subsec:Model-2:-Five}. Shaded regions in the plot
depict 95\% CL exclusions over the parameter space, with the exception
of the green (light green) region which is preferred by a global fit
to $b\rightarrow s\mu\mu$ data at 2$\sigma$ (3$\sigma$) \cite{Alguero:2023jeh}.
The dashed line represents the natural benchmark $g_{1}\simeq g_{2}\simeq g_{3}\simeq\sqrt{3}g_{Y}$
motivated in the main text. The dashed-dotted line represents the
contour where $\mathcal{B}(\tau\rightarrow3\mu)=10^{-12}$.\label{fig:Zp23_plot}}
\end{figure}

$Z-Z'_{23}$ mixing also shifts the couplings of the $Z$ boson to
fermions, leading to an important impact over $Z$-pole electroweak
precision observables (EWPOs) if $Z'_{23}$ lives at the TeV scale.
We find bounds coming from tests of $Z$ boson lepton universality
and flavour-violating $Z$ decays not to be competitive with the bound
from $\rho$. The electron asymmetry parameter $A_{e}$, which already
deviates from the SM by almost $2\sigma$ \cite{ALEPH:2005ab}, is
expected to deviate further in our model. Nevertheless, we expect
our model to improve the fit of $A_{b}^{\mathrm{FB}}$, which is in
tension with the SM prediction by more than $2\sigma$ \cite{ALEPH:2005ab}.
In conclusion, the global effect of our model over EWPOs can only
by captured by performing a global fit, which we leave for future
work. In this direction, a dedicated phenomenological analysis along
the lines of \cite{FernandezNavarro:2022gst,Navarro:2021sfb} would
be much needed. Global fits of EWPOs in the context of other $Z'$
models have been performed in the literature, see e.g.~\cite{Erler:2009ut,Allanach:2021kzj,Allanach:2022bik},
which obtain 95\% CL maximum values of $\mathrm{sin}\theta_{Z-Z'}$
ranging from $0.002$ to $0.0006$ depending on the model. We expect
our model to lie on the more restrictive side of that range. In principle,
our model could explain the anomalous CDF $M_{W}$ measurement, however
in this case we expect the contributions to other EWPOs to be intolerably
large, worsening the global fit.

The massive $Z'_{23}$ boson has sizable couplings to light quarks
and light charged leptons unless $g_{3}$ is very large, which we
do not expect based on naturalness arguments and also to protect the
extrapolation of the model in the UV, as discussed in Section~\ref{subsec:Couplings_Zprime}.
Consequently, on general grounds we expect a significant production
of a TeV-scale $Z'_{23}$ at the LHC, plus a sizable branching fraction
to electrons and muons. We have prepared the UFO model of $Z'_{23}$
using \texttt{FeynRules} \cite{Feynrules:2013bka}, and then we have
computed the $Z'_{23}$ production cross section for 13 TeV $pp$
collisions using \texttt{Madgraph5} \cite{Madgraph:2014hca} with
the default PDF \texttt{NNPDF23LO}. We estimated analytically the
branching fraction to electrons and muons, and we computed the total
decay width via the narrow width approximation. We confront our results
with the limits from the most recent dilepton resonance searches by
ATLAS \cite{ATLAS:2019erb} and CMS \cite{CMS:2021ctt} in order to
obtain 95\% CL exclusion bounds. The bounds from ditau \cite{ATLAS:2017eiz}
and ditop \cite{ATLAS:2020lks} searches turn out not to be competitive
even for the region of large $g_{3}$, where the bound from the $\rho$
electroweak parameter is stronger. Our results are depicted as the
black-shaded region in Fig.~\ref{fig:Zp23_plot}. As expected, the
bounds become weaker in the region $g_{3}>1$ where the couplings
to light fermions become mildly suppressed. In the region of small
$g_{3}$ we find the opposite behavior, such that LHC limits can exclude
$Z'_{23}$ as heavy as 6-7 TeV. After combining the LHC exclusion
with the bounds coming from the $\rho$ electroweak parameter, we
conclude that we can find $Z'_{23}$ as light as 3.5 TeV for $g_{3}=1$,
while for the benchmark $g_{i}\simeq\sqrt{3}g_{Y}$ we obtain $M_{Z'_{23}}\apprge5\;\mathrm{TeV}$.
\begin{table}
\centering{}%
\begin{tabular}{cc}
\toprule 
$Z'_{23}$ decay mode  & $\mathcal{B}$\tabularnewline
\midrule
\midrule 
$\bar{t}t$  & $\sim0.28$\tabularnewline
$\bar{u}u+\bar{c}c$  & $\sim0.14$\tabularnewline
$\bar{t}c+\bar{c}t$  & $\sim10^{-4}$\tabularnewline
\midrule 
$\bar{b}b$  & $\sim0.08$\tabularnewline
$\bar{d}d+\bar{s}s$  & $\sim0.04$\tabularnewline
$\bar{b}s+\bar{s}b$  & $\sim10^{-4}$\tabularnewline
\midrule 
$\tau^{+}\tau^{-}$  & $\sim0.25$\tabularnewline
$e^{+}e^{-}+\mu^{+}\mu^{-}$  & $\sim0.12$\tabularnewline
$\tau^{+}\mu^{-}+\tau^{-}\mu^{+}$  & $\sim10^{-5}$\tabularnewline
\midrule 
$\bar{\nu}\nu$  & $\sim0.08$\tabularnewline
\bottomrule
\end{tabular}\caption{Main decay modes of $Z'_{23}$ for the natural benchmark $g_{1}\simeq g_{2}\simeq g_{3}\simeq\sqrt{3}g_{Y}$.
We assume that decays into SM singlet neutrinos are kinematically
forbidden or suppressed.\textcolor{orange}{{} \label{tab:Main-decay-modes}}}
\end{table}

Given that $Z'_{23}$ has sizable couplings to electrons, we have
studied the bounds over contact interactions obtained at LEP \cite{Electroweak:2003ram}.
For our model, the most competitive bounds arise from contact interactions
involving only electrons. Assuming vector-like interactions, the bounds
by LEP are only sensitive to regions with very small $g_{3}$, but
can exclude $Z'_{23}$ masses beyond 10 TeV. However, we expect this
bound to be slightly overestimated for our model, since the interactions
of $Z'_{23}$ are not exactly vector-like due to the different hypercharge
of $e_{L}$ and $e_{R}$, as depicted in Eq.~\eqref{eq:Z23_couplings}.
Nevertheless, the bounds over chiral operators are much weaker than
the bound over the vector-like operator, and a dedicated reanalysis
of the data would be required in order to obtain the proper bound
for our model, which is beyond the scope of this work. Therefore,
we prefer to be conservative and depict the largest bound of the vector-like
operator as the purple region in Fig.~\ref{fig:Zp23_plot}.

We have also considered implications for $B$-physics. The heavy boson
$Z'_{23}$ has a sizable left-handed $b_{L}s_{L}$ coupling and an
approximately vector-like and universal coupling to electron and muon
pairs. Given these features, a $Z'_{23}$ with a mass of 2 TeV mediates
a meaningful contribution to the effective operator $\mathcal{O}_{9}^{\ell\ell}=(\bar{s}_{L}\gamma_{\mu}b_{L})(\bar{\ell}\gamma^{\mu}\ell)$
(with $\ell=e,\mu$), where sizable NP contributions are preferred
according to the most recent global fits \cite{Alguero:2023jeh},
without contributing to the SM-like $R_{K^{(*)}}$ ratios \cite{LHCb:2022qnv}.
However, as depicted in Fig.~\ref{fig:Zp23_plot}, the region where
the model could address the anomalies in $b\rightarrow s\mu\mu$ data
are in tension with the bounds obtained by dilepton searches, as expected
for a $Z'$ which has sizable couplings to light quarks. Nevertheless,
we can see that a relevant $C_{9}^{\ell\ell}\sim0.1$ can be obtained
for a heavier $Z'_{23}$ in the region where $g_{3}<0.5$, as in this
region couplings to light fermions are enhanced. 

$\mathcal{B}(B_{s}\rightarrow\mu\mu)$ is also enhanced above the
SM prediction due to both $Z'_{23}$ and $Z$ exchange diagrams. In
the region of small $g_{3}$ the $Z'_{23}$ exchange dominates, while
for large $g_{3}$ the $Z$ exchange dominates. As anticipated before,
the couplings of the $Z'_{23}$ boson to muons are approximately,
but not completely, vector-like. Therefore, a small contribution to
the operator $\mathcal{O}_{10}^{\mu\mu}=(\bar{s}_{L}\gamma_{\mu}b_{L})(\bar{\mu}\gamma^{\mu}\gamma_{5}\mu)$
is generated in the region of small $g_{3}$, where $Z'_{23}$ couplings
to muons are larger, leading to $Z'_{23}$ exchange being dominant.
However, in the region of large $g_{3}$, the flavour-violating coupling
$\bar{s}_{L}b_{L}Z'_{23}$ mixes into $\bar{s}_{L}b_{L}Z$ via $Z-Z'_{23}$
mixing. This provides an effective contribution to $\mathcal{O}_{10}^{\mu\mu}$
mediated by the $Z$ boson that enhances $\mathcal{B}(B_{s}\rightarrow\mu\mu)$
above the SM prediction. Overall, the region excluded at 95\% CL by
the current HFLAV average \cite{HFLAV:2022wzx} (including the latest
measurement by CMS \cite{CMS:2022mgd}) is depicted as blue-shaded
in Fig.~\ref{fig:Zp23_plot}. It is clear that the resulting bound
is not currently competitive with that from the $\rho$ electroweak
parameter which constrains the size of $Z-Z'_{23}$ mixing.

The flavour-violating structure of $Z'_{23}$ fermion couplings leads
to sizable contributions to $B_{s}-\bar{B}_{s}$ meson mixing \cite{DiLuzio:2019jyq}
and CLFV processes involving $\tau\rightarrow\mu$ transitions and
$\tau\rightarrow e$ transitions, although well below existing experimental
limits \cite{PDG:2022ynf}. $\tau\rightarrow\mu(e)$ transitions arise
from mixing angles connected to the flavour hierarchy $\mathcal{O}(m_{2}/m_{3})$
($\mathcal{O}(m_{1}/m_{3})$), see Section~\ref{subsec:Lessons-from-the}.
As an example, in Fig.~\ref{fig:Zp23_plot} we depict the contour
for $\mathcal{B}(\tau\rightarrow3\mu)=10^{-12}$. We find $\tau\rightarrow\mu\gamma$
to be more competitive than $\tau\rightarrow3\mu$ only in the region
$g_{3}>1$.

Beyond indirect detection, in the near future $Z'_{23}$ could be
directly produced at the LHC, HL-LHC and future colliders such as
FCC or a high energy muon collider. The particular pattern of $Z'_{23}$
fermion couplings will allow to disentangle our model from all other
proposals. For the natural benchmark $g_{1}\simeq g_{2}\simeq g_{3}\simeq\sqrt{3}g_{Y}$,
$Z'_{23}$ preferentially decays to top pairs and ditaus, as can be
seen in Table~\ref{tab:Main-decay-modes}. Furthermore, $Z'_{23}$
preferentially couples and decays to \textit{right-handed} charged
fermions, given their larger hypercharge with respect to left-handed
charged fermions. Similarly, decays to down-type quarks are generally
suppressed with respect to (right-handed) up-type quarks and charged
leptons, given the smaller hypercharge of the former. Due to the modification
of EWPOs, our model can also be tested in an electroweak precision
machine such as FCC-$ee$. An alternative way of discovery would be
the detection of the hyperon scalars breaking the $U(1)_{Y}^{3}$
group down to SM hypercharge, however we leave a study about the related
phenomenology for future work. In the same spirit, the model naturally
predicts SM singlet neutrinos which could be as light as the TeV scale,
with phenomenological implications yet to be explored in a future
work.

\section{Conclusions\label{sec:Conclusions}}

We have proposed a tri-hypercharge (TH) embedding of the Standard
Model, based on assigning a separate gauged weak hypercharge to each
family. The idea is that each fermion family $i$ only carries hypercharge
under a corresponding $U(1)_{Y_{i}}$ factor. This ensures that each
family transforms differently under the TH gauge group $U(1)_{Y}^{3}$,
which avoids the family repetition of the SM and provides the starting
point for a theory of flavour.

The three family specific hypercharge groups are spontaneously broken
in a cascade symmetry breaking down to SM hypercharge. We have motivated
a particular symmetry breaking pattern, where in a first step $U(1)_{Y_{1}}\times U(1)_{Y_{2}}$
is broken down to its diagonal subgroup at a high scale $v_{12}$.
The remaining group $U(1)_{Y_{1}+Y_{2}}\times U(1)_{Y_{3}}$ is broken
down to SM hypercharge at a scale $v_{23}$. This symmetry breaking
pattern sequentially recovers the accidental flavour symmetry of the
SM, providing protection versus FCNCs that allows the NP scales to
be relatively low. Dynamics connected to the scale $v_{12}$ play
a role in the origin of the family hierarchy $m_{1}/m_{2}$, while
dynamics connected to the scale $v_{23}$ play a role in the origin
of $m_{2}/m_{3}$. The hierarchy of scales $v_{23}/v_{12}$ generally
plays a role on the origin of flavour hierarchies as well, although
we have found that a mild hierarchy $v_{23}/v_{12}\simeq\lambda$
is enough for specific implementations of the model, where $\lambda\simeq0.224$
is the Wolfenstein parameter.

Assuming that the SM Higgs only carries third family hypercharge,
then only the third family Yukawa couplings are allowed at renormalisable
level. This explains the heaviness of the third family, the smallness
of $V_{cb}$ and $V_{ub}$ quark mixing, and delivers an accidental
$U(2)^{5}$ flavour symmetry in the Yukawa sector acting on the light
families, which provides a reasonable first order description of the
SM spectrum. However, $U(2)^{5}$ does not explain the hierarchical
heaviness of the top quark with respect to the bottom and tau fermions.
Furthermore, we have proven that the model generates a similar mass
hierarchy for all charged sectors, being unable to explain the heaviness
of the charm quark with respect to the strange and muon without small
couplings. We have motivated the addition of a second Higgs doublet
as a natural and elegant solution, which allows a more natural description
of the hierarchies between the different charged fermion sectors.

We have explored the capabilities of the $U(1)_{Y}^{3}$ model to
explain the observed hierarchies in the charged fermion sector, via
the addition of non-renormalisable operators containing $U(1)_{Y}^{3}$-breaking
scalars which act as small breaking effects of $U(2)^{5}$. After
extracting model-independent considerations from the spurion formalism,
we have presented example models where all charged fermion masses
and mixings are addressed. Following a similar methodology, we have
studied the origin of neutrino masses and mixing in the TH model.
We have shown that due to the $U(1)_{Y}^{3}$ gauge symmetry, the
implementation of a type I seesaw mechanism naturally leads to a low
scale seesaw, where the SM singlet neutrinos in the model may be as
light as the TeV scale. We have provided an example model compatible
with the observed pattern of neutrino mixing.

Finally, we have performed a preliminary exploration of the phenomenological
implications and discovery prospects of the $U(1)_{Y}^{3}$ theory
of flavour. The heavy gauge boson $Z'_{12}$ arising from the 12-breaking
displays completely flavour non-universal couplings to fermions, and
generally contributes to $\Delta F=2$ and CLFV processes. The size
of the most dangerous contributions are however model-dependent. In
selected specific models provided in this manuscript, we have found
that the most dangerous contributions to $K^{0}-\bar{K}^{0}$ mixing
and $\mu\rightarrow e\gamma$ are strongly suppressed, allowing for
$Z'_{12}$ to be as light as 10-50 TeV. Therefore, the lightest gauge
boson $Z'_{23}$ arising from the 23-breaking can live at the TeV
scale, within the reach of LHC and future colliders, since $Z'_{23}$
avoids bounds from $K^{0}-\bar{K}^{0}$ mixing and $\mu\rightarrow e\gamma$
thanks to an accidental GIM mechanism for light fermions.

We find the gauge boson $Z'_{23}$ to have a rich low energy phenomenology:
mixing with the SM $Z$ boson leads to implications for the $W$ boson
mass and EWPOs, plus we expect contributions to flavour-violating
processes involving the third family, such as $\tau\rightarrow3\mu(e)$
and $B_{s}-\bar{B}_{s}$ meson mixing. After our preliminary analysis,
we find that current data allows $Z'_{23}$ to be as light as 3-4
TeV in some regions of the parameter space, the strongest bounds coming
from dilepton searches at LHC along with the contribution to the $\rho$
electroweak parameter. In the case of discovery, the particular pattern
of $Z'_{23}$ couplings and decays to fermions will allow to disentangle
our model from all other proposals. However, most of the phenomenological
consequences are yet to be explored in detail: a global fit to EWPOs
and flavour observables will allow to properly confront our model
versus current data. An alternative way of discovery would be the
detection of the Higgs scalars (hyperons) breaking the $U(1)_{Y}^{3}$
down to SM hypercharge, however we leave a discussion about the related
phenomenology for future work. In the same spirit, the model naturally
predicts SM singlet neutrinos which could be as light as the TeV,
with phenomenological implications yet to be explored. The tri-hypercharge
gauge group may be the first step towards understanding the origin
of three fermion masses in Nature, the hierarchical charged fermion
masses and CKM mixing, revealing the existence of a flavour non-universal
gauge structure encoded in Nature at energies above the electroweak
scale.

\section*{Acknowledgements}

We would like to thank the CERN Theory group for hospitality and financial
support during an intermediate stage of this work. This work has received
funding from the European Union's Horizon 2020 Research and Innovation
Programme under Marie Sk\l odowska-Curie grant agreement HIDDeN European
ITN project (H2020-MSCA-ITN-2019//860881-HIDDeN). SFK acknowledges
the STFC Consolidated Grant ST/L000296/1.

\appendix

\section{General formalism for the seesaw mechanism\label{sec:General-formalism-for-seesaw}}

In Section~\ref{sec:Neutrino-masses-and-Mixing} we have motivated
that in order to explain the observed pattern of neutrino mixing via
a type I seesaw mechanism, it is required to add SM singlet neutrinos
which carry family hypercharges (but are SM singlets). Therefore,
the latter have to be vector-like in order to cancel gauge anomalies.
In all generality, such vector-like neutrinos can get contributions
to their mass from the unspecified vector-like mass terms and from
Majorana masses given by VEVs of hyperons breaking the $U(1)_{Y}^{3}$
group. Both the SM singlet neutrinos $N$ and the conjugate partners
$\overline{N}$ (both LH in our convention) can couple to the active
neutrinos $\nu$ through non-renormalisable operators involving the
hyperons. This way, one can write the general neutrino matrix, 
\begin{equation}
M_{\nu}=\left(
\global\long\def\arraystretch{0.7}%
\begin{array}{@{}llcc@{}}
 & \multicolumn{1}{c@{}}{\phantom{\!\,}\nu} & \phantom{\!\,}\overline{N} & \phantom{\!\,}N\\
\cmidrule(l){2-4}\left.\:\,\nu\right| & 0 & m_{D_{L}} & m_{D_{R}}\\
\left.\overline{N}\right| & m_{D_{L}}^{\mathrm{T}} & M_{L} & M_{LR}\\
\left.N\right| & m_{D_{R}}^{\mathrm{T}} & M_{LR}^{\mathrm{T}} & M_{R}
\end{array}\right)\equiv\left(\begin{array}{cc}
0 & m_{D}\\
m_{D}^{\mathrm{T}} & M_{N}
\end{array}\right)\,.\label{eq:Full_Mnu}
\end{equation}
In order to match Eq.~\eqref{eq:Full_Mnu} to our example seesaw
model of Section~\ref{subsec:Example-of-seesaw}, we define $\nu$
as a 3-component vector containing the weak eigenstates of active
neutrinos, while $N$ and $\overline{N}$ are 2-component vectors
containing the SM singlets $N$ and conjugate neutrinos $\overline{N}$
, respectively. Similarly, in the following we define each of the
sub-matrices of Eq.~\eqref{eq:Full_Mnu} in terms of the matter content
of Section~\ref{subsec:Example-of-seesaw}, ignoring the $\mathcal{O}(1)$
dimensionless couplings: 
\begin{flalign}
 &  &  & m_{D_{L}}=\left(
\global\long\def\arraystretch{0.7}%
\begin{array}{@{}llc@{}}
 & \multicolumn{1}{c@{}}{\phantom{\!\,}\overline{N}_{\mathrm{sol}}} & \phantom{\!\,}\overline{N}_{\mathrm{atm}}\\
\cmidrule(l){2-3}\left.L_{1}\right| & 0 & 0\\
\left.L_{2}\right| & 0 & 0\\
\left.L_{3}\right| & \frac{\tilde{\phi}_{\nu13}}{\Lambda_{\mathrm{sol}}} & \frac{\phi_{\mathrm{atm}}}{\Lambda_{\mathrm{atm}}}
\end{array}\right)H_{u}\,, &  & m_{D_{R}}=\left(
\global\long\def\arraystretch{0.7}%
\begin{array}{@{}llc@{}}
 & \multicolumn{1}{c@{}}{\phantom{\!\,}N_{\mathrm{sol}}} & \phantom{\!\,}N_{\mathrm{atm}}\\
\cmidrule(l){2-3}\left.L_{1}\right| & \frac{\phi_{e12}}{\Lambda_{\mathrm{sol}}} & 0\\
\left.L_{2}\right| & \frac{\tilde{\phi}_{e12}}{\Lambda_{\mathrm{sol}}} & \frac{\phi_{\mathrm{atm}}}{\Lambda_{\mathrm{atm}}}\\
\left.L_{3}\right| & \frac{\phi_{\nu13}}{\Lambda_{\mathrm{sol}}} & \frac{\tilde{\phi}_{\mathrm{atm}}}{\Lambda_{\mathrm{atm}}}
\end{array}\right)H_{u}\,,\label{eq:mDL_mDR}\\
\nonumber \\ &  &  & M_{L}=\left(
\global\long\def\arraystretch{0.7}%
\begin{array}{@{}llc@{}}
 & \multicolumn{1}{c@{}}{\phantom{\!\,}\overline{N}_{\mathrm{sol}}} & \phantom{\!\,}\overline{N}_{\mathrm{atm}}\\
\cmidrule(l){2-3}\left.\:\:\overline{N}_{\mathrm{sol}}\right| & \tilde{\phi}_{\mathrm{sol}} & 0\\
\left.\overline{N}_{\mathrm{atm}}\right| & 0 & \tilde{\phi}_{\ell23}
\end{array}\right)\approx v_{23}\mathbb{I}_{2\times2}\,, &  & M_{R}\approx\left(
\global\long\def\arraystretch{0.7}%
\begin{array}{@{}llc@{}}
 & \multicolumn{1}{c@{}}{\phantom{\!\,}N_{\mathrm{sol}}} & \phantom{\!\,}N_{\mathrm{atm}}\\
\cmidrule(l){2-3}\left.\:\:N_{\mathrm{sol}}\right| & \phi_{\mathrm{sol}} & 0\\
\left.N_{\mathrm{atm}}\right| & 0 & \phi_{\ell23}
\end{array}\right)\approx v_{23}\mathbb{I}_{2\times2}\,,\label{eq:ML_MR}
\end{flalign}
\begin{equation}
M_{LR}=\left(
\global\long\def\arraystretch{0.7}%
\begin{array}{@{}llc@{}}
 & \multicolumn{1}{c@{}}{\phantom{\!\,}N_{\mathrm{sol}}} & \phantom{\!\,}N_{\mathrm{atm}}\\
\cmidrule(l){2-3}\left.\:\:\overline{N}_{\mathrm{sol}}\right| & M_{N_{\mathrm{sol}}} & 0\\
\left.\overline{N}_{\mathrm{atm}}\right| & 0 & M_{N_{\mathrm{atm}}}
\end{array}\right)\approx M_{\mathrm{VL}}\mathbb{I}_{2\times2}\,,\label{eq:MLR}
\end{equation}
where $\mathbb{I}_{2\times2}$ is the $2\times2$ identity matrix.
In Eqs.~\eqref{eq:ML_MR} and \eqref{eq:MLR} above, we have considered
the following approximations and assumptions: 
\begin{itemize}
\item We have neglected $\mathcal{O}(1)$ dimensionless couplings generally
present for each non-zero entry of Eq.~\eqref{eq:ML_MR}. With this
consideration, we find $M_{L}=M_{R}$ after the hyperons develop their
VEVs. Furthermore, since the two hyperons appearing in $M_{L}$ and
$M_{R}$ participate in the 23-breaking step of Eq.~\eqref{eq:Symmetry_Breaking},
we have assumed that they both develop a similar VEV $\left\langle \phi_{\ell23}\right\rangle \approx\left\langle \phi_{\mathrm{sol}}\right\rangle \approx\mathcal{O}(v_{23})$.
For simplicity we take them to be equal, although the same conclusions
hold as long as they just differ by $\mathcal{O}(1)$ factors, as
naturally expected.
\item For simplicity, we have assumed a similar vector-like mass for both
neutrinos in Eq.~\eqref{eq:MLR}, i.e.~$M_{N_{\mathrm{sol}}}\approx M_{N_{\mathrm{atm}}}\equiv M_{\mathrm{VL}}$. 
\end{itemize}
Dirac-type masses in $m_{D_{L,R}}$ may be orders of magnitude smaller
than the electroweak scale, because they arise from non-renormalisable
operators proportional to the SM VEV. In contrast, the eigenvalues
of $M_{N}$ are not smaller than $\mathcal{O}(v_{23})$, which is
at least TeV. Therefore, the condition $m_{D}\ll M_{N}$ is fulfilled
in Eq.~\eqref{eq:Full_Mnu} and we can safely apply the seesaw formula
as 
\begin{align}
m_{\nu} & \simeq m_{D}M_{N}^{-1}m_{D}^{\mathrm{T}}\label{eq:Seesaw_General}\\
 & =\left(\begin{array}{cc}
m_{D_{L}} & m_{D_{R}}\end{array}\right)\left(\begin{array}{cc}
v_{23} & -M_{\mathrm{VL}}\\
-M_{\mathrm{VL}} & v_{23}
\end{array}\right)\left(\begin{array}{c}
m_{D_{L}}^{\mathrm{T}}\\
m_{D_{R}}^{\mathrm{T}}
\end{array}\right)\frac{1}{v_{23}^{2}-M_{\mathrm{VL}}^{2}}\nonumber \\
 & =\left[m_{D_{L}}m_{D_{L}}^{\mathrm{T}}v_{23}-m_{D_{L}}m_{D_{R}}^{\mathrm{T}}M_{\mathrm{VL}}-m_{D_{R}}m_{D_{L}}^{\mathrm{T}}M_{\mathrm{VL}}+m_{D_{R}}m_{D_{R}}^{\mathrm{T}}v_{23}\right]\frac{1}{v_{23}^{2}-M_{\mathrm{VL}}^{2}}\,.\nonumber 
\end{align}
Given the structure of $m_{D_{L}}$ and $m_{D_{R}}$ in Eq.~(\ref{eq:mDL_mDR}),
the products above involving $m_{D_{L}}$ lead to a hierarchical $m_{\nu}$
matrix where only the entries in the third row and column are populated
and the others are zero. In contrast, the product $m_{D_{R}}m_{D_{R}}^{\mathrm{T}}$
provides a matrix $m_{\nu}$ where all entries are populated. Therefore,
if $M_{\mathrm{VL}}\gg v_{23}$, then the effective neutrino matrix
becomes hierarchical, rendering impossible to explain the observed
pattern of neutrino mixing and mass splittings with $\mathcal{O}(1)$
parameters. Instead, if $M_{\mathrm{VL}}$ is of the same order or
smaller than $v_{23}$, i.e.~$M_{\mathrm{VL}}\apprle v_{23}$, then
the resulting matrix is in any case a matrix where all entries are
populated, which has the potential to explain the observed patterns
of neutrino mixing. As shown in the main text, Section~\ref{subsec:Example-of-seesaw},
by further assuming the mild hierarchies $v_{23}/v_{12}\simeq\lambda$
and $\Lambda_{\mathrm{sol}}/\Lambda_{\mathrm{atm}}\simeq\lambda$
the product $m_{D_{R}}m_{D_{R}}^{\mathrm{T}}$ above provides a texture
for $m_{\nu}$ that can reproduce the observed neutrino oscillation
data with $\mathcal{O}(1)$ parameters.

This argument holds as long as $m_{D_{L}}$ is populated by zeros
in at least some of the entries involving $L_{1}$ and $L_{2}$, like
in our example model of Section~\ref{subsec:Example-of-seesaw}.
Instead, in the very particular case where both $m_{D_{L}}$ and $m_{D_{R}}$
are similarly populated matrices and of the same order, then the terms
proportional to $M_{\mathrm{VL}}$ in Eq.~\eqref{eq:Seesaw_General}
can provide an effective neutrino mass matrix with $\mathcal{O}(1)$
coefficients in each entry. In this case, $M_{\mathrm{VL}}>v_{23}$
is possible. Nevertheless, even in this scenario we expect $M_{\mathrm{VL}}$
not to be very large, since the smallness of $m_{D_{L}}$ and $m_{D_{R}}$
(that arise from non-renormalisable operators) may potentially provide
most of the suppression for the small neutrino masses. Furthermore,
this scenario involves the addition of several extra hyperons with
very particular charges, making the model more complicated, so we
do not consider it.

\section{High scale symmetry breaking\label{sec:High-scale-symmetry}}

Assuming that the 12-breaking scale is far above the electroweak scale,
at very high energies we consider only the factors $U(1)_{Y_{1}}\times U(1)_{Y_{2}}$
with renormalisable Lagrangian (neglecting fermion content and any
kinetic mixing\footnote{Considering kinetic mixing in the Lagrangian of Eq.~\eqref{eq:Lagrangian_12}
only leads to a redefinition of either the $g_{1}$ or $g_{2}$ couplings
in the canonical basis (where the kinetic terms are diagonal).} for simplicity),
\begin{align}
\mathcal{L} & =-\frac{1}{4}F_{\mu\nu}^{(1)}F^{\mu\nu(1)}-\frac{1}{4}F_{\mu\nu}^{(2)}F^{\mu\nu(2)}+(D_{\mu}\phi_{12})^{*}D^{\mu}\phi_{12}-V(\phi_{12})\,,\label{eq:Lagrangian_12}
\end{align}
where for simplicity we assume only one hyperon $\phi_{12}(q,-q)$,
which develops a VEV $\left\langle \phi_{12}\right\rangle =v_{12}/\sqrt{2}$
spontaneously breaking $U(1)_{Y_{1}}\times U(1)_{Y_{2}}$ down to
its diagonal subgroup. The covariant derivative reads
\begin{equation}
D_{\mu}=\partial_{\mu}-ig_{1}Y_{1}B_{1\mu}-ig_{2}Y_{2}B_{2\mu}\,.
\end{equation}
Expanding the kinetic term of $\phi_{12}$, we obtain mass terms for
the gauge bosons as
\begin{equation}
{\displaystyle \mathcal{M}^{2}=\frac{q^{2}v_{12}^{2}}{2}\left(
\global\long\def\arraystretch{0.7}%
\begin{array}{@{}lcc@{}}
 & \phantom{\!\,}B{}_{1}^{\mu} & \phantom{\!\,}B{}_{2}^{\mu}\\
\cmidrule(l){2-3}\left.B_{1\mu}\right| & g_{1}^{2} & -g_{1}g_{2}\\
\left.B_{2\mu}\right| & -g_{1}g_{2} & g_{2}^{2}
\end{array}\right)\,.}
\end{equation}
The diagonalisation of the matrix above reveals
\begin{equation}
\hat{\mathcal{M}}^{2}=\frac{q^{2}v_{12}^{2}}{2}\left(
\global\long\def\arraystretch{0.7}%
\begin{array}{@{}lcc@{}}
 & \phantom{\!\,}Y{}_{12}^{\mu} & \phantom{\!\,}Z'{}_{12}^{\mu}\\
\cmidrule(l){2-3}\left.\;\,Y_{12\mu}\right| & 0 & 0\\
\left.\,\,Z'_{12\mu}\right| & 0 & g_{1}^{2}+g_{2}^{2}
\end{array}\right)\,,\label{eq:MassMatrixGauge_Simple}
\end{equation}
in the basis of mass eigenstates given by
\begin{equation}
\left(\begin{array}{c}
Y_{12\mu}\\
Z'_{12\mu}
\end{array}\right)=\left(\begin{array}{cc}
\cos\theta_{12} & \sin\theta_{12}\\
-\sin\theta_{12} & \cos\theta_{12}
\end{array}\right)\left(\begin{array}{c}
B_{1\mu}\\
B_{2\mu}
\end{array}\right)\,,\qquad\sin\theta_{12}=\frac{g_{1}}{\sqrt{g_{1}^{2}+g_{2}^{2}}}\,.
\end{equation}
Therefore, we obtain a massive gauge boson $Z'_{12\mu}$ at the scale
$v_{12}$, while $Y_{1}+Y_{2}$ associated to the gauge boson $Y_{12\mu}$
remains unbroken. These results are trivially generalised for the
case of more hyperons. The covariant derivative in the new basis is
given by
\begin{align}
D_{\mu} & =\partial_{\mu}-i\frac{g_{1}g_{2}}{\sqrt{g_{1}^{2}+g_{2}^{2}}}(Y_{1}+Y_{2})Y_{12\mu}-i\left(-\frac{g_{1}^{2}}{\sqrt{g_{1}^{2}+g_{2}^{2}}}Y_{1}+\frac{g_{2}^{2}}{\sqrt{g_{1}^{2}+g^{2}}}Y_{2}\right)Z'_{12\mu}\label{eq:Covariant_Derivative_12}\\
 & =\partial_{\mu}-ig_{12}(Y_{1}+Y_{2})Y_{12\mu}-i\left(-g_{1}\sin\theta_{12}Y_{1}+g_{2}\cos\theta_{12}Y_{2}\right)Z'_{12\mu}\,.\nonumber 
\end{align}
The fermion couplings in Eq.~\eqref{eq:Z12_couplings} are readily
extracted by expanding the fermion kinetic terms applying Eq.~\eqref{eq:Covariant_Derivative_12}.

\section{Low scale symmetry breaking\label{sec:Low-scale-symmetry}}

The renormalisable Lagrangian of a theory $SU(2)_{L}\times U(1)_{Y_{1}+Y_{2}}\times U(1)_{Y_{3}}$
with $H(\mathbf{2})_{(0,\frac{1}{2})}$ and $\phi_{23}(\mathbf{1})_{(q,-q)}$
reads (neglecting fermion content and kinetic mixing, although we
will consider the effect of kinetic mixing at the end of the section),
\begin{align}
\mathcal{L}_{\mathrm{ren}} & =-\frac{1}{4}F_{\mu\nu}^{(12)}F^{\mu\nu(12)}-\frac{1}{4}F_{\mu\nu}^{(3)}F^{\mu\nu(3)}-\frac{1}{4}W_{\mu\nu}^{a}W_{a}^{\mu\nu}\label{eq:23_breaking_Lagrangian}\\
 & +(D_{\mu}H)^{\dagger}D^{\mu}H+(D_{\mu}\phi_{23})^{*}D^{\mu}\phi_{23}\nonumber \\
 & -V(H,\phi_{23})\,,\nonumber 
\end{align}
where the covariant derivatives read
\begin{equation}
D_{\mu}H=(\partial_{\mu}-ig_{L}\frac{\sigma^{a}}{2}W_{\mu}^{a}-i\frac{g_{3}}{2}B_{3\mu})H\,,\label{eq:cov1}
\end{equation}
\begin{equation}
D_{\mu}\phi_{23}=(\partial_{\mu}-ig_{12}qB_{12\mu}+ig_{3}qB_{3\mu})\phi_{23}\,,\label{eq:cov2}
\end{equation}
and $\sigma^{a}$ with $a=1,2,3$ are the Pauli matrices. The Higgs
doublet develops the usual electroweak symmetry breaking VEV as
\begin{equation}
\left\langle H\right\rangle =\frac{1}{\sqrt{2}}\left(\begin{array}{c}
0\\
v_{\mathrm{SM}}
\end{array}\right)\,,
\end{equation}
while the hyperon develops a higher scale VEV as
\begin{equation}
\left\langle \phi_{23}\right\rangle =\frac{v_{23}}{\sqrt{2}}\,,
\end{equation}
which spontaneously breaks the group $U(1)_{Y_{1}+Y_{2}}\times U(1)_{Y_{3}}$
down to its diagonal subgroup.

Expanding the following kinetic terms with the expressions of the
covariant derivatives of Eqs.~\eqref{eq:cov1} and \eqref{eq:cov2},
we obtain
\begin{equation}
(D_{\mu}H)^{\dagger}D^{\mu}H+(D_{\mu}\phi_{23})^{*}D^{\mu}\phi_{23}=\frac{v_{\mathrm{SM}}^{2}g_{L}^{2}}{4}W_{\mu}W^{\mu\dagger}+\frac{q^{2}v_{23}^{2}}{2}\left(
\global\long\def\arraystretch{0.7}%
\begin{array}{@{}llcc@{}}
 & \multicolumn{1}{c@{}}{\phantom{\!\,}W_{3}^{\mu}} & \phantom{\!\,}B_{12}^{\mu} & \phantom{\!\,}B_{3}^{\mu}\\
\cmidrule(l){2-4}\left.\,W_{3\mu}\right| & g_{L}^{2}r^{2} & 0 & -g_{L}g_{3}r^{2}\\
\left.B_{12\mu}\right| & 0 & g_{12}^{2} & -g_{12}g_{3}\\
\left.\;\,B_{3\mu}\right| & -g_{L}g_{3}r^{2} & -g_{12}g_{3} & g_{3}^{2}+g_{3}^{2}r^{2}
\end{array}\right),
\end{equation}
where $r=\frac{v_{\mathrm{SM}}}{2qv_{23}}\ll1$, we have defined $W_{\mu}=(W_{\mu}^{1}+iW_{\mu}^{2})/\sqrt{2}$,
and we denote $M_{\mathrm{gauge}}^{2}$ as the off-diagonal matrix
above. Given the two different scales in the mass matrix above, we
first apply the following transformation 
\begin{equation}
\left(\begin{array}{c}
W_{3}^{\mu}\\
Y^{\mu}\\
X^{\mu}
\end{array}\right)=\left(\begin{array}{ccc}
1 & 0 & 0\\
0 & \cos\theta_{23} & \sin\theta_{23}\\
0 & -\sin\theta_{23} & \cos\theta_{23}
\end{array}\right)\left(\begin{array}{c}
W_{3}^{\mu}\\
B_{12}^{\mu}\\
B_{3}^{\mu}
\end{array}\right)=\left(\begin{array}{c}
W_{3}^{\mu}\\
\cos\theta_{23}B_{12}^{\mu}+\sin\theta_{23}B_{3}^{\mu}\\
-\sin\theta_{23}B_{12}^{\mu}+\cos\theta_{23}B_{3}^{\mu}
\end{array}\right)\,,\label{eq:C.7}
\end{equation}
where 
\begin{equation}
\sin\theta_{23}=\frac{g_{12}}{\sqrt{g_{12}^{2}+g_{3}^{2}}}\,,
\end{equation}
and we denote the rotation in Eq.~\eqref{eq:C.7} as $V_{\theta_{23}}$,
obtaining 
\begin{equation}
V_{\theta_{23}}M_{\mathrm{gauge}}^{2}V_{\theta_{23}}^{\dagger}=\frac{q^{2}v_{23}^{2}}{2}\left(
\global\long\def\arraystretch{0.7}%
\begin{array}{@{}llcc@{}}
 & \multicolumn{1}{c@{}}{\phantom{\!\,}W_{3}^{\mu}} & \phantom{\!\,}Y^{\mu} & \phantom{\!\,}X^{\mu}\\
\cmidrule(l){2-4}\left.W_{3\mu}\right| & g_{L}^{2}r^{2} & -g_{L}g_{Y}r^{2} & -g_{L}g_{X}r^{2}\\
\left.\;\,\,\,Y_{\mu}\right| & -g_{L}g_{Y}r^{2} & g_{Y}^{2}r^{2} & g_{Y}g_{X}r^{2}\\
\left.\;\;X_{\mu}\right| & -g_{L}g_{X}r^{2} & g_{Y}g_{X}r^{2} & g_{F}^{2}+g_{X}^{2}r^{2}
\end{array}\right)\,,\label{eq:MassMatrix_MatchLiterature}
\end{equation}
where $Y^{\mu}$ is the SM hypercharge gauge boson with gauge coupling
\begin{equation}
g_{Y}=\frac{g_{12}g_{3}}{\sqrt{g_{12}^{2}+g_{3}^{2}}}\simeq0.36\,,
\end{equation}
where the numeric value depicted is evaluated at the electroweak scale,
and $X^{\mu}$ can be interpreted as an effective gauge boson with
effective couplings 
\begin{equation}
g_{X}=\frac{g_{3}^{2}}{\sqrt{g_{12}^{2}+g_{3}^{2}}}\,,\qquad\qquad g_{F}=\sqrt{g_{12}^{2}+g_{3}^{2}}\,,
\end{equation}
to the Higgs boson and to $\phi_{23}$, respectively. In this basis,
the covariant derivatives read 
\begin{equation}
D_{\mu}H=(\partial_{\mu}-ig_{L}\frac{\sigma^{a}}{2}W_{\mu}^{a}-i\frac{g_{Y}}{2}Y_{\mu}-i\frac{g_{X}}{2}X_{\mu})H\,,
\end{equation}
\begin{equation}
D_{\mu}\phi_{23}=(\partial_{\mu}-iqg_{F}X_{\mu})\phi_{23}\,.
\end{equation}
The mass matrix in this basis can be block-diagonalised by applying
the following transformation 
\begin{equation}
\left(\begin{array}{c}
A^{\mu}\\
(Z^{0})^{\mu}\\
X^{\mu}
\end{array}\right)=\left(\begin{array}{ccc}
\sin\theta_{W} & \cos\theta_{W} & 0\\
\cos\theta_{W} & -\sin\theta_{W} & 0\\
0 & 0 & 1
\end{array}\right)\left(\begin{array}{c}
W_{3}^{\mu}\\
Y^{\mu}\\
X^{\mu}
\end{array}\right)=\left(\begin{array}{c}
\cos\theta_{W}Y^{\mu}+\sin\theta_{W}W_{3}^{\mu}\\
-\sin\theta_{W}Y^{\mu}+\cos\theta_{W}W_{3}^{\mu}\\
X^{\mu}
\end{array}\right)\,,\label{eq:Basis_BeforeZ-ZprimeMixing}
\end{equation}
where the mixing angle is identified with the usual weak mixing angle
as 
\begin{equation}
\sin\theta_{W}=\frac{g_{Y}}{\sqrt{g_{Y}^{2}+g_{L}^{2}}}\,,
\end{equation}
and we denote the matrix in Eq.~\eqref{eq:Basis_BeforeZ-ZprimeMixing}
as $V_{\theta_{W}}$\footnote{Notice that this is not the usual SM convention, because we have ordered
$W_{\mu}^{3}$ and $Y_{\mu}$ differently.}, obtaining 
\begin{equation}
V_{\theta_{W}}V_{\theta_{23}}M_{\mathrm{gauge}}^{2}(V_{\theta_{W}}V_{\theta_{23}})^{\dagger}=\frac{q^{2}v_{23}^{2}}{2}\left(
\global\long\def\arraystretch{0.7}%
\begin{array}{@{}llcc@{}}
 & \multicolumn{1}{c@{}}{\phantom{\!\,}A^{\mu}} & \phantom{\!\,}(Z^{0})^{\mu} & \phantom{\!\,}X^{\mu}\\
\cmidrule(l){2-4}\left.\quad\;A_{\mu}\right| & 0 & 0 & 0\\
\left.(Z^{0})_{\mu}\right| & 0 & (g_{L}^{2}+g_{Y}^{2})r^{2} & -g_{X}\sqrt{g_{Y}^{2}+g_{L}^{2}}r^{2}\\
\left.\quad\,X_{\mu}\right| & 0 & -g_{X}\sqrt{g_{Y}^{2}+g_{L}^{2}}r^{2} & g_{F}^{2}+g_{X}^{2}r^{2}
\end{array}\right)\,,
\end{equation}
where we have already identified the massless photon. Now we diagonalise
the remaining $2\times2$ sub-block in the limit of small $r^{2}$.
We obtain 
\begin{equation}
Z_{\mu}=\cos\theta_{Z-Z'_{23}}\left(-\sin\theta_{W}Y_{\mu}+\cos\theta_{W}W_{3\mu}\right)+\sin\theta_{Z-Z'_{23}}X_{\mu}\,,
\end{equation}
\begin{equation}
Z'_{23\mu}=-\sin\theta_{Z-Z'_{23}}\left(-\sin\theta_{W}Y_{\mu}+\cos\theta_{W}W_{3\mu}\right)+\cos\theta_{Z-Z'_{23}}X_{\mu}\,,
\end{equation}
where to leading order in $r^{2}$ 
\begin{equation}
\sin\theta_{Z-Z'_{23}}\approx\frac{\sqrt{g_{Y}^{2}+g_{L}^{2}}g_{X}}{g_{F}^{2}}r^{2}=\frac{g_{3}\cos\theta_{23}}{\sqrt{g_{Y}^{2}+g_{L}^{2}}}\left(\frac{M_{Z}^{0}}{M_{Z'_{23}}^{0}}\right)^{2}=\frac{\sqrt{g_{3}^{2}-g_{Y}^{2}}}{\sqrt{g_{Y}^{2}+g_{L}^{2}}}\left(\frac{M_{Z}^{0}}{M_{Z'_{23}}^{0}}\right)^{2}\,,\label{eq:Z_Zp_mixing}
\end{equation}
where we have used the matching condition with SM hypercharge to write
everything in terms of $g_{3}$ and $g_{Y}$. We can see that the
SM $Z$ boson carries a small admixture of the $X_{\mu}$ boson, which
provides a small shift to its mass as 
\begin{equation}
M_{Z}^{2}\approx q^{2}v_{23}^{2}\left(g_{Y}^{2}+g_{L}^{2}\right)\left(r^{2}-\frac{g_{X}^{2}}{g_{F}^{2}}r^{4}\right)=(M_{Z}^{0})^{2}\left[1-\frac{g_{3}^{2}-g_{Y}^{2}}{g_{Y}^{2}+g_{L}^{2}}\left(\frac{M_{Z}^{0}}{M_{Z'_{23}}^{0}}\right)^{2}\right]\,,
\end{equation}
\begin{equation}
M_{Z'_{23}}^{2}\approx q^{2}v_{23}^{2}g_{F}^{2}\left(1+\frac{g_{X}^{2}}{g_{F}^{2}}r^{2}\right)=(M_{Z'_{23}}^{0})^{2}\left[1+\frac{g_{3}^{2}-g_{Y}^{2}}{g_{Y}^{2}+g_{L}^{2}}\left(\frac{M_{Z}^{0}}{M_{Z'_{23}}^{0}}\right)^{2}\right]\,,
\end{equation}
where 
\begin{equation}
M_{Z}^{0}=\frac{v_{\mathrm{SM}}}{2}\sqrt{g_{Y}^{2}+g_{L}^{2}}\,,\qquad\qquad M_{Z'_{23}}^{0}=qv_{\mathrm{23}}\sqrt{g_{12}^{2}+g_{3}^{2}}=qv_{\mathrm{23}}\frac{g_{3}^{2}}{\sqrt{g_{3}^{2}-g_{Y}^{2}}},
\end{equation}
are the masses of the $Z$ boson in the SM and the mass of the $Z'_{23}$
boson in absence of $Z-Z'_{23}$ mixing, respectively. All these results
can be easily generalised for the case of more hyperons or more Higgs
doublets.

As expected, the SM $Z$ boson mass arises at order $r^{2}$, with
a leading correction from $Z-Z'_{23}$ mixing arising at order $r^{4}$.
Instead, the $Z'_{23}$ boson arises at leading order in the power
expansion, with the leading correction from $Z-Z'_{23}$ mixing arising
at order $r^{2}$. Remarkably, the presence of $Z-Z'_{23}$ mixing
always shifts the mass of the $Z$ boson to smaller values with respect
to the SM prediction.

The equations obtained match general results in the literature \cite{Allanach:2018lvl,Langacker:2008yv,Bandyopadhyay:2018cwu},
which consider scenarios where the starting point is a matrix such
as Eq.~\eqref{eq:MassMatrix_MatchLiterature} with $g_{F}=g_{X}$.
Our equations match those of these papers when $g_{F}=g_{X}$ (and
taking into account that we need to perform an extra rotation $\theta_{23}$
to arrive to Eq.~\eqref{eq:MassMatrix_MatchLiterature}). 

In the case that a kinetic mixing term $\sin\chi F_{\mu\nu}^{(12)}F_{\mu\nu}^{(3)}/2$
is included in Eq.~\eqref{eq:23_breaking_Lagrangian}, then one can
repeat the calculations of this section to finally obtain
\begin{equation}
\sin\theta_{Z-Z'_{23}}=\frac{g_{3}\cos\theta_{23}}{\sqrt{g_{Y}^{2}+g_{L}^{2}}}(1-\sin\chi)\left(\frac{M_{Z}^{0}}{M_{Z'_{23}}^{0}}\right)^{2}\,,\label{eq:Z_Zp_mixing_kinetic}
\end{equation}
where now
\begin{equation}
\cos\theta_{23}=\frac{g_{3}\sec^{2}\chi}{\sqrt{g_{12}^{2}+g_{3}^{2}\left(\sec\chi-\tan\chi\right)^{2}}}\,,\qquad g_{Y}=\frac{g_{12}g_{3}\sec^{2}\chi}{\sqrt{g_{12}^{2}+g_{3}^{2}\left(\sec\chi-\tan\chi\right)^{2}}}\,,
\end{equation}
\begin{equation}
M_{Z'_{23}}^{0}=qv_{\mathrm{23}}\sqrt{g_{12}^{2}+g_{3}^{2}\left(\frac{1-\sin\chi}{1+\sin\chi}\right)}\,.
\end{equation}
Assuming that the kinetic mixing parameter $\sin\chi$ is small compared
to unity (e.g.~absent at tree-level and generated by loop diagrams),
then the dominant effect is always the original gauge mixing and kinetic
mixing can be neglected. As an example, the hyperon $\phi_{23}(\mathbf{1})_{(q,-q)}$
charged under both $U(1)$ groups generates kinetic mixing at 1-loop
as $d\sin\chi/d\log\mu=-g_{12}g_{3}q^{2}/(16\pi^{2})$, which leads
to $\sin\chi(\mu)=g_{12}g_{3}q^{2}\log(m_{\phi_{23}}^{2}/\mu^{2})/(16\pi^{2})$.
For the natural benchmark $g_{12}\approx\sqrt{3/2}g_{Y}$ and $g_{3}\approx\sqrt{3}g_{Y}$
motivated in Section~\ref{subsec:Couplings_Zprime}, along with typical
values $q=1/2$ and $m_{\phi_{23}}=1\,\mathrm{TeV}$, we obtain $\sin\chi(M_{Z})\simeq0.002$.

Neglecting the small $Z-Z'_{23}$ mixing, the fermion couplings of
the $Z'_{23}$ gauge boson given in Eq.~\eqref{eq:Z23_couplings}
are obtained by expanding the fermion kinetic terms in the usual way,
using the covariant derivative (where $T_{3}$ is the third-component
$SU(2)_{L}$ isospin, and we do not include the terms associated to
charge currents nor QCD interactions) 
\begin{flalign}
D_{\mu} & =\partial_{\mu}-i\left[eQA_{\mu}+\left(T_{3}g_{L}\cos\theta_{W}-g_{Y}\sin\theta_{W}(Y_{1}+Y_{2}+Y_{3})\right)Z_{\mu}^{0}\right.\label{eq:cov_23}\\
 & \left.+\left(-g_{12}\sin\theta_{23}(Y_{1}+Y_{2})+g_{3}\cos\theta_{23}Y_{3}\right)Z'_{23\mu}\right]\nonumber \\
 & =D_{\mu}^{\mathrm{SM}}-i\left(-\frac{g_{Y}^{2}}{\sqrt{g_{3}^{2}-g_{Y}^{2}}}(Y_{1}+Y_{2})+\sqrt{g_{3}^{2}-g_{Y}^{2}}Y_{3}\right)Z'_{23\mu}\,,
\end{flalign}
which is an excellent approximation for all practical purposes other
than precision $Z$ boson phenomenology. In that case, one has to
consider that the couplings of the $Z$ boson to fermions are shifted
due to $Z-Z'_{23}$ mixing as
\begin{equation}
g_{Z}^{f_{L}f_{L}}=\left(g_{Z}^{f_{L}f_{L}}\right)^{0}+\sin\theta_{Z-Z'_{23}}g_{Z'_{23}}^{f_{L}f_{L}}\,,
\end{equation}
where $g_{Z'_{23}}^{f_{L}f_{L}}$ are the fermion couplings of $Z'_{23}$
in the absence of $Z-Z'_{23}$ mixing, as given in Eq.~\eqref{eq:Z23_couplings},
and similarly for right-handed fermions by just replacing $L$ by
$R$ everywhere. We can see that in any case, the shift in the $Z$
boson couplings is suppressed by the small ratio $(M_{Z}^{0}/M_{Z'_{23}}^{0})^{2}$.

\providecommand{\href}[2]{#2}\begingroup\raggedright\endgroup


\begin{thebibliography}{10}

\bibitem{PDG:2022ynf}
{\scshape Particle Data Group} collaboration, R.~L. Workman, \emph{{Review of
  Particle Physics}}, {\emph{PTEP} {\bfseries 2022} (2022) 083C01}.

\bibitem{deSalas:2020pgw}
P.~F. de~Salas, D.~V. Forero, S.~Gariazzo, P.~Mart\'\i{}nez-Mirav\'e, O.~Mena,
  C.~A. Ternes et~al., \emph{{2020 global reassessment of the neutrino
  oscillation picture}},
  \href{https://doi.org/10.1007/JHEP02(2021)071}{\emph{JHEP} {\bfseries 02}
  (2021) 071} [\href{https://arxiv.org/abs/2006.11237}{{\ttfamily
  2006.11237}}].

\bibitem{Gonzalez-Garcia:2021dve}
M.~C. Gonzalez-Garcia, M.~Maltoni and T.~Schwetz, \emph{{NuFIT: Three-Flavour
  Global Analyses of Neutrino Oscillation Experiments}},
  \href{https://doi.org/10.3390/universe7120459}{\emph{Universe} {\bfseries 7}
  (2021) 459} [\href{https://arxiv.org/abs/2111.03086}{{\ttfamily
  2111.03086}}].

\bibitem{Li:1981nk}
X.~Li and E.~Ma, \emph{{Gauge Model of Generation Nonuniversality}},
  \href{https://doi.org/10.1103/PhysRevLett.47.1788}{\emph{Phys. Rev. Lett.}
  {\bfseries 47} (1981) 1788}.

\bibitem{Ma:1987ds}
E.~Ma, X.~Li and S.~F. Tuan, \emph{{Gauge Model of Generation Nonuniversality
  Revisited}}, \href{https://doi.org/10.1103/PhysRevLett.60.495}{\emph{Phys.
  Rev. Lett.} {\bfseries 60} (1988) 495}.

\bibitem{Ma:1988dn}
E.~Ma and D.~Ng, \emph{{Gauge and Higgs Bosons in a Model of Generation
  Nonuniversality}}, \href{https://doi.org/10.1103/PhysRevD.38.304}{\emph{Phys.
  Rev. D} {\bfseries 38} (1988) 304}.

\bibitem{Li:1992fi}
X.-y. Li and E.~Ma, \emph{{Gauge model of generation nonuniversality
  reexamined}}, \href{https://doi.org/10.1088/0954-3899/19/9/006}{\emph{J.
  Phys. G} {\bfseries 19} (1993) 1265}
  [\href{https://arxiv.org/abs/hep-ph/9208210}{{\ttfamily hep-ph/9208210}}].

\bibitem{Hill:1994hp}
C.~T. Hill, \emph{{Topcolor assisted technicolor}},
  \href{https://doi.org/10.1016/0370-2693(94)01660-5}{\emph{Phys. Lett. B}
  {\bfseries 345} (1995) 483}
  [\href{https://arxiv.org/abs/hep-ph/9411426}{{\ttfamily hep-ph/9411426}}].

\bibitem{Muller:1996dj}
D.~J. Muller and S.~Nandi, \emph{{Top flavor: A Separate SU(2) for the third
  family}}, \href{https://doi.org/10.1016/0370-2693(96)00745-9}{\emph{Phys.
  Lett. B} {\bfseries 383} (1996) 345}
  [\href{https://arxiv.org/abs/hep-ph/9602390}{{\ttfamily hep-ph/9602390}}].

\bibitem{Malkawi:1996fs}
E.~Malkawi, T.~M.~P. Tait and C.~P. Yuan, \emph{{A Model of strong flavor
  dynamics for the top quark}},
  \href{https://doi.org/10.1016/0370-2693(96)00859-3}{\emph{Phys. Lett. B}
  {\bfseries 385} (1996) 304}
  [\href{https://arxiv.org/abs/hep-ph/9603349}{{\ttfamily hep-ph/9603349}}].

\bibitem{Craig:2011yk}
N.~Craig, D.~Green and A.~Katz, \emph{{(De)Constructing a Natural and Flavorful
  Supersymmetric Standard Model}},
  \href{https://doi.org/10.1007/JHEP07(2011)045}{\emph{JHEP} {\bfseries 07}
  (2011) 045} [\href{https://arxiv.org/abs/1103.3708}{{\ttfamily 1103.3708}}].

\bibitem{Panico:2016ull}
G.~Panico and A.~Pomarol, \emph{{Flavor hierarchies from dynamical scales}},
  \href{https://doi.org/10.1007/JHEP07(2016)097}{\emph{JHEP} {\bfseries 07}
  (2016) 097} [\href{https://arxiv.org/abs/1603.06609}{{\ttfamily
  1603.06609}}].

\bibitem{Barbieri:2021wrc}
R.~Barbieri, \emph{{A View of Flavour Physics in 2021}},
  \href{https://doi.org/10.5506/APhysPolB.52.789}{\emph{Acta Phys. Polon. B}
  {\bfseries 52} (2021) 789}
  [\href{https://arxiv.org/abs/2103.15635}{{\ttfamily 2103.15635}}].

\bibitem{Bordone:2017bld}
M.~Bordone, C.~Cornella, J.~Fuentes-Martin and G.~Isidori, \emph{{A three-site
  gauge model for flavor hierarchies and flavor anomalies}},
  \href{https://doi.org/10.1016/j.physletb.2018.02.011}{\emph{Phys. Lett. B}
  {\bfseries 779} (2018) 317}
  [\href{https://arxiv.org/abs/1712.01368}{{\ttfamily 1712.01368}}].

\bibitem{Allwicher:2020esa}
L.~Allwicher, G.~Isidori and A.~E. Thomsen, \emph{{Stability of the Higgs
  Sector in a Flavor-Inspired Multi-Scale Model}},
  \href{https://doi.org/10.1007/JHEP01(2021)191}{\emph{JHEP} {\bfseries 01}
  (2021) 191} [\href{https://arxiv.org/abs/2011.01946}{{\ttfamily
  2011.01946}}].

\bibitem{Fuentes-Martin:2020pww}
J.~Fuentes-Martin, G.~Isidori, J.~Pag\`es and B.~A. Stefanek, \emph{{Flavor
  non-universal Pati-Salam unification and neutrino masses}},
  \href{https://doi.org/10.1016/j.physletb.2021.136484}{\emph{Phys. Lett. B}
  {\bfseries 820} (2021) 136484}
  [\href{https://arxiv.org/abs/2012.10492}{{\ttfamily 2012.10492}}].

\bibitem{Fuentes-Martin:2022xnb}
J.~Fuentes-Martin, G.~Isidori, J.~M. Lizana, N.~Selimovic and B.~A. Stefanek,
  \emph{{Flavor hierarchies, flavor anomalies, and Higgs mass from a warped
  extra dimension}},
  \href{https://doi.org/10.1016/j.physletb.2022.137382}{\emph{Phys. Lett. B}
  {\bfseries 834} (2022) 137382}
  [\href{https://arxiv.org/abs/2203.01952}{{\ttfamily 2203.01952}}].

\bibitem{Davighi:2022bqf}
J.~Davighi, G.~Isidori and M.~Pesut, \emph{{Electroweak-flavour and
  quark-lepton unification: a family non-universal path}},
  \href{https://doi.org/10.1007/JHEP04(2023)030}{\emph{JHEP} {\bfseries 04}
  (2023) 030} [\href{https://arxiv.org/abs/2212.06163}{{\ttfamily
  2212.06163}}].

\bibitem{Davighi:2022fer}
J.~Davighi and J.~Tooby-Smith, \emph{{Electroweak flavour unification}},
  \href{https://doi.org/10.1007/JHEP09(2022)193}{\emph{JHEP} {\bfseries 09}
  (2022) 193} [\href{https://arxiv.org/abs/2201.07245}{{\ttfamily
  2201.07245}}].

\bibitem{King:2021jeo}
S.~F. King, \emph{{Twin Pati-Salam theory of flavour with a TeV scale vector
  leptoquark}}, \href{https://doi.org/10.1007/JHEP11(2021)161}{\emph{JHEP}
  {\bfseries 11} (2021) 161}
  [\href{https://arxiv.org/abs/2106.03876}{{\ttfamily 2106.03876}}].

\bibitem{FernandezNavarro:2022gst}
M.~Fern\'andez~Navarro and S.~F. King, \emph{{B-anomalies in a twin Pati-Salam
  theory of flavour including the 2022 LHCb $ {R}_{K^{\left(\ast \right)}} $
  analysis}}, \href{https://doi.org/10.1007/JHEP02(2023)188}{\emph{JHEP}
  {\bfseries 02} (2023) 188}
  [\href{https://arxiv.org/abs/2209.00276}{{\ttfamily 2209.00276}}].

\bibitem{FernandezNavarro:2023lgk}
M.~Fern\'andez~Navarro, \emph{{Flavour hierarchies and $B$-anomalies in a twin
  Pati-Salam theory of flavour}},  in \emph{{8th Symposium on Prospects in the
  Physics of Discrete Symmetries}}, 2, 2023,
  \href{https://arxiv.org/abs/2302.10829}{{\ttfamily 2302.10829}}.

\bibitem{FernandezNavarro:2023ykw}
M.~Fern\'andez~Navarro, \emph{{Twin Pati-Salam theory of flavour for a new
  picture of $B$-anomalies}},  in \emph{{57th Rencontres de Moriond on
  Electroweak Interactions and Unified Theories}}, 5, 2023,
  \href{https://arxiv.org/abs/2305.19356}{{\ttfamily 2305.19356}}.

\bibitem{Ferretti:2006df}
L.~Ferretti, S.~F. King and A.~Romanino, \emph{{Flavour from accidental
  symmetries}},
  \href{https://doi.org/10.1088/1126-6708/2006/11/078}{\emph{JHEP} {\bfseries
  11} (2006) 078} [\href{https://arxiv.org/abs/hep-ph/0609047}{{\ttfamily
  hep-ph/0609047}}].

\bibitem{Barbieri:2011ci}
R.~Barbieri, G.~Isidori, J.~Jones-Perez, P.~Lodone and D.~M. Straub,
  \emph{{$U(2)$ and Minimal Flavour Violation in Supersymmetry}},
  \href{https://doi.org/10.1140/epjc/s10052-011-1725-z}{\emph{Eur. Phys. J. C}
  {\bfseries 71} (2011) 1725}
  [\href{https://arxiv.org/abs/1105.2296}{{\ttfamily 1105.2296}}].

\bibitem{Davighi:2023iks}
J.~Davighi and G.~Isidori, \emph{{Non-universal gauge interactions addressing
  the inescapable link between Higgs and Flavour}},
  \href{https://arxiv.org/abs/2303.01520}{{\ttfamily 2303.01520}}.

\bibitem{Allanach:2018lvl}
B.~C. Allanach and J.~Davighi, \emph{{Third family hypercharge model for $
  {R}_{K^{\left(\ast \right)}} $ and aspects of the fermion mass problem}},
  \href{https://doi.org/10.1007/JHEP12(2018)075}{\emph{JHEP} {\bfseries 12}
  (2018) 075} [\href{https://arxiv.org/abs/1809.01158}{{\ttfamily
  1809.01158}}].

\bibitem{Beniwal:2022kyv}
A.~Beniwal, F.~Rajec, M.~T. Prim, P.~Scott, W.~Su, M.~White et~al.,
  \emph{{Global fit of 2HDM with future collider results}},  in \emph{{Snowmass
  2021}}, 3, 2022, \href{https://arxiv.org/abs/2203.07883}{{\ttfamily
  2203.07883}}.

\bibitem{deGiorgi:2023wjh}
A.~de~Giorgi, F.~Koutroulis, L.~Merlo and S.~Pokorski, \emph{{Flavour and Higgs
  physics in $Z_2$-symmetric 2HD models near the decoupling limit}},
  \href{https://arxiv.org/abs/2304.10560}{{\ttfamily 2304.10560}}.

\bibitem{UTfit:2007eik}
{\scshape UTfit} collaboration, M.~Bona et~al., \emph{{Model-independent
  constraints on $\Delta F=2$ operators and the scale of new physics}},
  \href{https://doi.org/10.1088/1126-6708/2008/03/049}{\emph{JHEP} {\bfseries
  03} (2008) 049} [\href{https://arxiv.org/abs/0707.0636}{{\ttfamily
  0707.0636}}].

\bibitem{Isidori:2014rba}
G.~Isidori and F.~Teubert, \emph{{Status of indirect searches for New Physics
  with heavy flavour decays after the initial LHC run}},
  \href{https://doi.org/10.1140/epjp/i2014-14040-4}{\emph{Eur. Phys. J. Plus}
  {\bfseries 129} (2014) 40} [\href{https://arxiv.org/abs/1402.2844}{{\ttfamily
  1402.2844}}].

\bibitem{King:1998jw}
S.~F. King, \emph{{Atmospheric and solar neutrinos with a heavy singlet}},
  \href{https://doi.org/10.1016/S0370-2693(98)01055-7}{\emph{Phys. Lett. B}
  {\bfseries 439} (1998) 350}
  [\href{https://arxiv.org/abs/hep-ph/9806440}{{\ttfamily hep-ph/9806440}}].

\bibitem{King:1999mb}
S.~F. King, \emph{{Large mixing angle MSW and atmospheric neutrinos from single
  right-handed neutrino dominance and U(1) family symmetry}},
  \href{https://doi.org/10.1016/S0550-3213(00)00109-7}{\emph{Nucl. Phys. B}
  {\bfseries 576} (2000) 85}
  [\href{https://arxiv.org/abs/hep-ph/9912492}{{\ttfamily hep-ph/9912492}}].

\bibitem{King:2002nf}
S.~F. King, \emph{{Constructing the large mixing angle MNS matrix in seesaw
  models with right-handed neutrino dominance}},
  \href{https://doi.org/10.1088/1126-6708/2002/09/011}{\emph{JHEP} {\bfseries
  09} (2002) 011} [\href{https://arxiv.org/abs/hep-ph/0204360}{{\ttfamily
  hep-ph/0204360}}].

\bibitem{CDF:2022hxs}
{\scshape CDF} collaboration, T.~Aaltonen et~al., \emph{{High-precision
  measurement of the $W$ boson mass with the CDF II detector}},
  \href{https://doi.org/10.1126/science.abk1781}{\emph{Science} {\bfseries 376}
  (2022) 170}.

\bibitem{ATLAS:2023fsi}
{\scshape ATLAS} collaboration, \emph{{Improved W boson Mass Measurement using
  7 TeV Proton-Proton Collisions with the ATLAS Detector}}, .

\bibitem{ATLAS:2017rzl}
{\scshape ATLAS} collaboration, M.~Aaboud et~al., \emph{{Measurement of the
  $W$-boson mass in pp collisions at $\sqrt{s}=7$ TeV with the ATLAS
  detector}}, \href{https://doi.org/10.1140/epjc/s10052-017-5475-4}{\emph{Eur.
  Phys. J. C} {\bfseries 78} (2018) 110}
  [\href{https://arxiv.org/abs/1701.07240}{{\ttfamily 1701.07240}}].

\bibitem{Alguero:2023jeh}
M.~Alguer\'o, A.~Biswas, B.~Capdevila, S.~Descotes-Genon, J.~Matias and
  M.~Novoa-Brunet, \emph{{To (b)e or not to (b)e: No electrons at LHCb}},
  \href{https://arxiv.org/abs/2304.07330}{{\ttfamily 2304.07330}}.

\bibitem{ALEPH:2005ab}
{\scshape ALEPH, DELPHI, L3, OPAL, SLD, LEP Electroweak Working Group, SLD
  Electroweak Group, SLD Heavy Flavour Group} collaboration, S.~Schael et~al.,
  \emph{{Precision electroweak measurements on the $Z$ resonance}},
  \href{https://doi.org/10.1016/j.physrep.2005.12.006}{\emph{Phys. Rept.}
  {\bfseries 427} (2006) 257}
  [\href{https://arxiv.org/abs/hep-ex/0509008}{{\ttfamily hep-ex/0509008}}].

\bibitem{Navarro:2021sfb}
M.~Fern\'andez~Navarro and S.~F. King, \emph{{Fermiophobic $Z'$ model for
  simultaneously explaining the muon anomalies $R_{K^{(*)}}$ and
  $(g-2)_{\mu}$}},
  \href{https://doi.org/10.1103/PhysRevD.105.035015}{\emph{Phys. Rev. D}
  {\bfseries 105} (2022) 035015}
  [\href{https://arxiv.org/abs/2109.08729}{{\ttfamily 2109.08729}}].

\bibitem{Erler:2009ut}
J.~Erler, P.~Langacker, S.~Munir and E.~Rojas, \emph{{Constraints on the mass
  and mixing of Z-prime bosons}},
  \href{https://doi.org/10.1063/1.3327731}{\emph{AIP Conf. Proc.} {\bfseries
  1200} (2010) 790} [\href{https://arxiv.org/abs/0910.0269}{{\ttfamily
  0910.0269}}].

\bibitem{Allanach:2021kzj}
B.~C. Allanach, J.~E. Camargo-Molina and J.~Davighi, \emph{{Global fits of
  third family hypercharge models to neutral current B-anomalies and
  electroweak precision observables}},
  \href{https://doi.org/10.1140/epjc/s10052-021-09377-1}{\emph{Eur. Phys. J. C}
  {\bfseries 81} (2021) 721}
  [\href{https://arxiv.org/abs/2103.12056}{{\ttfamily 2103.12056}}].

\bibitem{Allanach:2022bik}
B.~Allanach and J.~Davighi, \emph{{$M_W$ helps select $Z^\prime $ models for
  $b\rightarrow s \ell \ell $ anomalies}},
  \href{https://doi.org/10.1140/epjc/s10052-022-10693-3}{\emph{Eur. Phys. J. C}
  {\bfseries 82} (2022) 745}
  [\href{https://arxiv.org/abs/2205.12252}{{\ttfamily 2205.12252}}].

\bibitem{Feynrules:2013bka}
A.~Alloul, N.~D. Christensen, C.~Degrande, C.~Duhr and B.~Fuks,
  \emph{{FeynRules 2.0 - A complete toolbox for tree-level phenomenology}},
  \href{https://doi.org/10.1016/j.cpc.2014.04.012}{\emph{Comput. Phys. Commun.}
  {\bfseries 185} (2014) 2250}
  [\href{https://arxiv.org/abs/1310.1921}{{\ttfamily 1310.1921}}].

\bibitem{Madgraph:2014hca}
J.~Alwall, R.~Frederix, S.~Frixione, V.~Hirschi, F.~Maltoni, O.~Mattelaer
  et~al., \emph{{The automated computation of tree-level and next-to-leading
  order differential cross sections, and their matching to parton shower
  simulations}}, \href{https://doi.org/10.1007/JHEP07(2014)079}{\emph{JHEP}
  {\bfseries 07} (2014) 079} [\href{https://arxiv.org/abs/1405.0301}{{\ttfamily
  1405.0301}}].

\bibitem{ATLAS:2019erb}
{\scshape ATLAS} collaboration, G.~Aad et~al., \emph{{Search for high-mass
  dilepton resonances using 139 fb$^{-1}$ of $pp$ collision data collected at
  $\sqrt{s}=$13 TeV with the ATLAS detector}},
  \href{https://doi.org/10.1016/j.physletb.2019.07.016}{\emph{Phys. Lett. B}
  {\bfseries 796} (2019) 68}
  [\href{https://arxiv.org/abs/1903.06248}{{\ttfamily 1903.06248}}].

\bibitem{CMS:2021ctt}
{\scshape CMS} collaboration, A.~M. Sirunyan et~al., \emph{{Search for resonant
  and nonresonant new phenomena in high-mass dilepton final states at $
  \sqrt{s} $ = 13 TeV}},
  \href{https://doi.org/10.1007/JHEP07(2021)208}{\emph{JHEP} {\bfseries 07}
  (2021) 208} [\href{https://arxiv.org/abs/2103.02708}{{\ttfamily
  2103.02708}}].

\bibitem{ATLAS:2017eiz}
{\scshape ATLAS} collaboration, M.~Aaboud et~al., \emph{{Search for additional
  heavy neutral Higgs and gauge bosons in the ditau final state produced in 36
  fb$^{-1}$ of pp collisions at $ \sqrt{s}=13 $ TeV with the ATLAS detector}},
  \href{https://doi.org/10.1007/JHEP01(2018)055}{\emph{JHEP} {\bfseries 01}
  (2018) 055} [\href{https://arxiv.org/abs/1709.07242}{{\ttfamily
  1709.07242}}].

\bibitem{ATLAS:2020lks}
{\scshape ATLAS} collaboration, G.~Aad et~al., \emph{{Search for $
  t\overline{t} $ resonances in fully hadronic final states in $pp$ collisions
  at $ \sqrt{s} $ = 13 TeV with the ATLAS detector}},
  \href{https://doi.org/10.1007/JHEP10(2020)061}{\emph{JHEP} {\bfseries 10}
  (2020) 061} [\href{https://arxiv.org/abs/2005.05138}{{\ttfamily
  2005.05138}}].

\bibitem{Electroweak:2003ram}
{\scshape LEP, ALEPH, DELPHI, L3, OPAL, LEP Electroweak Working Group, SLD
  Electroweak Group, SLD Heavy Flavor Group} collaboration, t.~S. Electroweak,
  \emph{{A Combination of preliminary electroweak measurements and constraints
  on the standard model}},
  \href{https://arxiv.org/abs/hep-ex/0312023}{{\ttfamily hep-ex/0312023}}.

\bibitem{LHCb:2022qnv}
{\scshape LHCb} collaboration, \emph{{Test of lepton universality in $b
  \rightarrow s \ell^+ \ell^-$ decays}},
  \href{https://arxiv.org/abs/2212.09152}{{\ttfamily 2212.09152}}.

\bibitem{HFLAV:2022wzx}
{\scshape Heavy Flavor Averaging Group, HFLAV} collaboration, Y.~S. Amhis
  et~al., \emph{{Averages of b-hadron, c-hadron, and \ensuremath{\tau}-lepton
  properties as of 2021}},
  \href{https://doi.org/10.1103/PhysRevD.107.052008}{\emph{Phys. Rev. D}
  {\bfseries 107} (2023) 052008}
  [\href{https://arxiv.org/abs/2206.07501}{{\ttfamily 2206.07501}}].

\bibitem{CMS:2022mgd}
{\scshape CMS} collaboration, A.~Tumasyan et~al., \emph{{Measurement of the
  B$^0_\mathrm{S}$$\to$$\mu^+\mu^-$ decay properties and search for the
  B$^0$$\to$$\mu^+\mu^-$ decay in proton-proton collisions at $\sqrt{s}$ = 13
  TeV}}, \href{https://doi.org/10.1016/j.physletb.2023.137955}{\emph{Phys.
  Lett. B} {\bfseries 842} (2023) 137955}
  [\href{https://arxiv.org/abs/2212.10311}{{\ttfamily 2212.10311}}].

\bibitem{DiLuzio:2019jyq}
L.~Di~Luzio, M.~Kirk, A.~Lenz and T.~Rauh, \emph{{$\Delta M_s$ theory precision
  confronts flavour anomalies}},
  \href{https://doi.org/10.1007/JHEP12(2019)009}{\emph{JHEP} {\bfseries 12}
  (2019) 009} [\href{https://arxiv.org/abs/1909.11087}{{\ttfamily
  1909.11087}}].

\bibitem{Langacker:2008yv}
P.~Langacker, \emph{{The Physics of Heavy $Z^\prime$ Gauge Bosons}},
  \href{https://doi.org/10.1103/RevModPhys.81.1199}{\emph{Rev. Mod. Phys.}
  {\bfseries 81} (2009) 1199}
  [\href{https://arxiv.org/abs/0801.1345}{{\ttfamily 0801.1345}}].

\bibitem{Bandyopadhyay:2018cwu}
T.~Bandyopadhyay, G.~Bhattacharyya, D.~Das and A.~Raychaudhuri,
  \emph{{Reappraisal of constraints on $Z^\prime$ models from unitarity and
  direct searches at the LHC}},
  \href{https://doi.org/10.1103/PhysRevD.98.035027}{\emph{Phys. Rev. D}
  {\bfseries 98} (2018) 035027}
  [\href{https://arxiv.org/abs/1803.07989}{{\ttfamily 1803.07989}}].

\end{thebibliography}
\end{document}